\documentclass[%
nofootinbib,
 aip,
 amsmath,amssymb,
 reprint,
 longbibliography,
]{revtex4-1}

\usepackage{graphicx}
\usepackage{dcolumn}
\usepackage{bm}

\usepackage[utf8]{inputenc}
\usepackage[T1]{fontenc}
\usepackage{mathptmx}
\usepackage{etoolbox}
\usepackage{comment}
\usepackage{amsfonts}
\usepackage{amsmath}
\usepackage{stackengine}
\usepackage{wrapfig}
\usepackage[x11names]{xcolor}
\usepackage{vwcol}  
\usepackage{array}
\usepackage{adjustbox}
\usepackage{multirow}
\usepackage[bottom]{footmisc}

\def\REM#1{} 
\def\N#1{#1}    
\def\simon#1{#1}

\makeatletter
\def\@email#1#2{%
 \endgroup
 \patchcmd{\titleblock@produce}
  {\frontmatter@RRAPformat}
  {\frontmatter@RRAPformat{\produce@RRAP{*#1\href{mailto:#2}{#2}}}\frontmatter@RRAPformat}
  {}{}
}%
\makeatother

\begin{document}

\preprint{AIP/123-QED}


\title{High-speed imaging of magnetized plasmas: when electron temperature matters}
\author{Simon Vincent}
\author{Vincent Dolique}
\author{Nicolas Plihon}


\affiliation{Univ Lyon, ENS de Lyon, CNRS, Laboratoire de Physique, F-69342 Lyon, France}

\date{\today}
          
\begin{abstract}
 High speed camera imaging is a powerful tool to probe the spatiotemporal features of unsteady processes in plasmas, usually assuming light fluctuations to be a proxy for the plasma density fluctuations. In this article, we systematically compare high speed camera imaging with simultaneous measurements of the plasma parameters -- plasma density, electron temperature, floating potential -- in a modestly magnetized Argon plasma column at low pressure (1 mTorr, magnetic fields ranging from 160 to 640~G). The light emission was filtered around $488\pm5$~nm, $750\pm5$~nm, $810\pm5$~nm. We show that the light intensity cannot be interpreted as a proxy for the plasma density and that the electron temperature cannot be ignored when interpreting high speed imaging, both for the time-averaged profiles and for the fluctuations. The features of plasma parameter fluctuations are investigated, with a focus on ion acoustic waves (at frequency around 70 kHz) at low magnetic field and low-frequency azimuthal waves (around a few kHz) at larger magnetic fields. An excellent match is found between the high speed images fluctuations and an Arrhenius law functional form which incorporates fluctuations of the plasma density and of the electron temperature. These results explain the discrepancies between ion saturation current and narrow-band imaging measurements previously reported in the literature.
\end{abstract}


\maketitle

\section{Introduction}

High-speed camera imaging is a useful tool to deduce time and space-resolved features of laboratory plasma dynamics from the light it naturally emits. Following the recent technological evolutions of high-speed cameras over the last two decades, a number of studies followed the path initiated decades ago in the fusion community~\cite{Carruthers_1957,Medley_1985}. For instance, recent investigations using high-speed visible imaging of tokamak plasmas allowed to gain insight into the dynamics of avaloids or blobs~\cite{Fuchert_2014, Tynan_2009}, filaments~\cite{Farley_2019}, the plasma response to magnetic perturbations~\cite{Angelini_2015}  or edge plasma dynamics~\cite{Cavalier_2019}. Indirect measurements of the local plasma dynamics Gas Puff Imaging~\cite{Zweben_2017} also offered great insights into the spatio-temporal dynamics of fusion plasmas. High speed imaging of lower temperature plasmas either dedicated to basic plasma studies or applications has experienced a recent boom. In the context of plasma propulsion, the transverse and streamwise dynamics of breathing modes~\cite{Desangles_2020} and rotating spokes~\cite{Mazouffre_2019} in Hall thrusters or MHD modes in hollow cathodes~\cite{Becatti_2021} benefited from images acquired up to several hundreds of thousands frames per seconds. In the context of basic plasma physics, the characterization of the dynamics of magnetized plasma archs benefited from high speed imaging~\cite{Hsu_2003, Soltwisch_2010, Sklodowski_2021}.  High speed imaging in linear plasma devices allow to gain insight on the ion dynamics~\cite{Claire_2018},  the transition to drift wave turbulence~\cite{Thakur_2014_1} or the generation of zonal flows~\cite{Xu_2019}. It can also be used for the comparison with simulations~\cite{Russel_2011}, or as a complement to other diagnostics~\cite{Thakur_2014_2, Brandt_2016}.
However, since the light radiated from a plasma depends in a very non-trivial way on the plasma parameters, extracting physical information from camera imaging is seldom very quantitative. However, camera imaging can be used alone as a conclusive diagnostic technique. In cylindrical geometry for instance, where the spatial fluctuations of an entire azimuthal wave can be captured at once, and if the propagation velocity is low enough, modern cameras allow the realization of detailed mode decompositions~\cite{Brandt_2011,Ohdachi_2017,Yamasaki_2017}. High speed plasma imaging has also been used for the extraction of experimental dispersion relations~\cite{Light_2013,Cui_2016} or the estimation of turbulent transport~\cite{Cui_2015}. In the latter examples, the light intensity is usually considered as a proxy for the ion saturation current, or for the density, with correlations up to 0.75~\cite{Cui_2015}. In low temperature and weakly ionized plasma, this approximation is usually by the Corona model~\cite{Fantz_2006,Cal_2010, Gonzalez_2020} which assumes that the main source for photons emission is radiative decay from excited states that are populated by electronic collisions from the ground state. As a consequence the intensity of the light emitted by the radiative decay from the electronic level $p$ to level $k$ of a species $\alpha$ satisfies:
\begin{equation}
    I_{pk} \propto n_e n_\alpha K_{ex}^{g \rightarrow p}(T_e),
\end{equation}
where $n_{e}$ is the electron density,  $n_\alpha$ is the ground state density of species $\alpha$,  and $K_{ex}^{g \rightarrow p}$ the excitation rate constant of state $p$ from the ground state of $\alpha$.
The densities of the neutrals and the electron temperature are usually assumed to be spatially homogeneous, and their temporal fluctuations are neglected. The emitted light then depends only on the plasma density: for light emitted by excited states of neutral species, the light intensity  $I_{\rm neutral} \propto n_e$ and for light emitted by excited states of ions in a single ion species plasma, the light intensity  $I_{\rm ion} \propto n_e^2$. While extremely convenient for the interpretation of camera images, this simplification is often crude, which is the reason for camera imaging to be mostly used as a qualitative diagnostic tool.

In the recent years great efforts were made to investigate the correlations between high speed camera imaging and electrostatic probes measurements  -- either the ion saturation current for an estimation of the density, or the floating potential -- in laboratory Argon plasmas~\cite{Antar_2007,Oldenburger_2010,Light_2013,Light_2019}. These studies usually showed a high correlation between  the ion saturation current and the broadband light fluctuations (typically spanning from 0.45~\cite{Antar_2007} to 0.6~\cite{Oldenburger_2010} or even 0.75~\cite{Cui_2015}), while it was significantly reduced when considering filtered light (typically reduced to the order of 0.4~\cite{Oldenburger_2010,Light_2019}).

In this article we present simultaneous measurements of plasma density, electron temperature and floating potential using an electrostatic probe, with high speed imaging of the plasma. To our knowledge, it is the first experimental study taking into account a well resolved spatial and temporal evolution of the electron temperature in order to interpret light intensity of visible light spectral lines directly recorded by camera in a low pressure, modestly magnetized plasma column. We show that, for typical conditions where low frequency waves are excited, the correlation between the light fluctuations and the ion saturation current is of the order of 0.7, and increases up to 0.87 for the electron temperature. The correlation with a model based on the Corona approximation and taking into account both the electron temperature and the plasma density slightly increases the correlation and reduces the time-lag with the light fluctuations. The experimental set-up and the electrostatic probe are first introduced in Section~\ref{sec:probecamera}. The high speed imaging optical system is then described: it consists of a high speed camera recording the light emitted by the plasma and filtered either around 488 nm (ArII line), 750 nm or 810 nm (ArI lines), and optical and geometrical corrections are taken into account for the image analysis. Based on excitation cross sections extracted from the bibliography, a model is proposed in Section~\ref{sec:Te} to take into account the dependence of high speed images with the electron temperature. Then using the time-averaged plasma parameter profiles, this model is compared to the light intensity radial profiles in Section~\ref{sec:mean}, and shows that spatial variations of the electron temperature cannot be ignored. The correlations between the instantaneous plasma parameters and the simultaneously recorded light intensity are then explored in Section~\ref{sec:fluctuations}, highlighting that the fluctuations of electron temperature is a key physical parameter to understand the temporal fluctuations of radiated light intensity. Finally, a first expansion of the proposed model, which clarifies the comparison between the camera recordings and the probe measurements, is discussed in Section~\ref{sec:discussion}. Conclusions are drawn in Section~\ref{sec:conclusion}.

\section{Experimental setup, electrostatic and optical diagnostics}\label{sec:probecamera}

\subsection{Experimental set-up}
\label{subsec:set-up}

\begin{figure}
    \centering
    \includegraphics[width = 0.98\columnwidth, trim={0in 0in 0in 0in},clip]{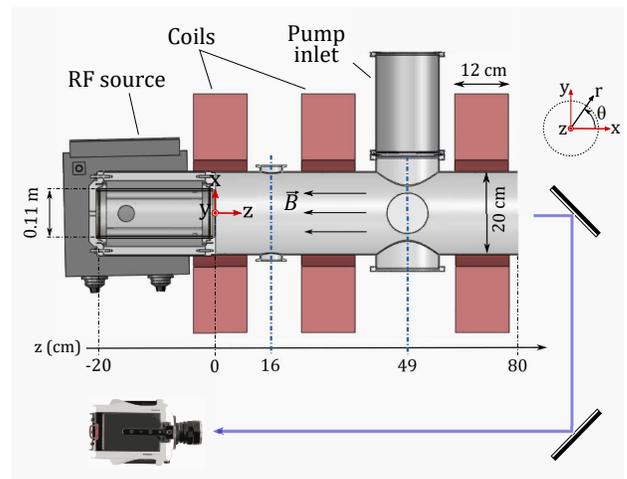}
    \caption{Sketch of the experimental set-up. Locations of probes are shown as dot-dashed lines. See text for details.}
    \label{vkp_sketch}
\end{figure}

The experimental setup consists of a magnetized plasma column, described in details elsewhere~\cite{Plihon_2015} and sketched in Fig.~\ref{vkp_sketch}. The plasma source is an inductive 3-turns helicoidal coil wrapped around a $20$~cm long, 11~cm inner diameter borosilicate tube, fed by a  3.3~kW, 13.56~MHz radio-frequency power supply through a manual L-type matching box. The source tube ends in a 11~cm diameter BN disk. The plasma expands in a  $20$~cm diameter, $80$~cm long grounded stainless steel cylindrical chamber, ending in a DN 200 borosilicate optical window, from which the light emitted by the plasma emitted is recorded by a high-speed camera. Three coils located along the chamber create a confining axial magnetic field. The magnetic field is not perfectly homogeneous along the $z$ direction and the values referred in the remaining of this article are taken  as the mean values along the axis, ranging from $160$~G to $640$~G. The inhomogeneity of the magnetic field has to be taken into account when comparing probe measurements and camera imaging; geometrical corrections are introduced in Subsection~\ref{subsec:camera_corrections}. Electrostatic probes can be inserted along two diameters along the $x$ axis, at positions $L_1 = 16$~cm and $L_2 = 49$~cm along $z$ (the origin $z=0$ taken at the junction between the source and the chamber). All results presented in this article were obtained for 1~kW of radio frequency power operated in Argon gas at a pressure of 1~mTorr.
To keep the whole apparatus in a steady thermal state, the operation of the plasma is pulsed: the plasma is sustained over typically 5 seconds, during which data are acquired, with a repetition period of typically 30~s. The experiment is fully automated to allow high repeatability and reproducibility of the plasma. The level of shot to shot reproducibility was $\pm 0.6\%$ for the ion saturation current of a Langmuir probe, with a standard deviation of $0.2\%$ (estimated from a series of 40 shots). Diametrical scans of the plasma parameters measured by the probe where acquired sequentially: each spatial point has been acquired during one plasma-pulse, and the probe is translated between two pulses.

\subsection{Probe measurements}
\label{sec:probes}

\begin{figure}[h!]
    \centering
    \includegraphics[width = 0.8\columnwidth, trim={0in 0in 0in 0in},clip]{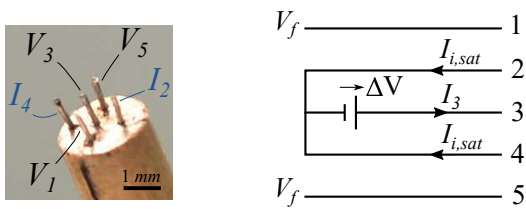}
    \caption{Left : photograph of the five-tips probe used for the measure of $I_i$, $V_f$ and $T_e$. Right : electrical sketch of the probe.}
    \label{ftips_sketch}
\end{figure}

Since the goal of the present article is to show that electron temperature fluctuations cannot be ignored when analyzing high speed camera imaging, we used a probe giving access to the instantaneous values of the electron temperature, plasma density and floating potential, know as a five tips probe~\cite{Tsui_1992,Theiler_2011}, very similar to a triple probe~\cite{Chen_1965}. A photograph of the probe and an electrical sketch are provided in Fig.~\ref{ftips_sketch}. \REM{The probe head is a 3~mm diameter BN cylinder from which five 1~mm long, 0.2~mm in diameter Tungsten wires spaced by 0.8~mm measure currents and voltages.} \N{The probe consists of five 1~mm long, 0.2~mm in diameter Tungsten wires spaced by 0.8~mm, housed in a 3~mm diameter BN cylinder.}
The plasma density $n$ \N{(assuming electroneutrality $n = n_e = n_i$, where $n_e$ is the electron density and $n_i$ the ion density)}, electron temperature $T_e$ and floating potential $V_f$ are then computed from currents and voltages \N{measured independently from the five tips}. This \REM{type of} probe also provides a measure of the turbulent transport, \REM{that is} not \N{detailed} \REM{presented} in this work. The raw electrical signals were conditioned using a home-made integrated electronics (current/voltage conversion \REM{converted using and} \N{and galvanic isolation using} opto-couplers) and  were digitized using a PXI-e 6368 data acquisition module at a frame rate of 200 kHz. \REM{Two tips (namely 1 and 5) measure the floating potential, from which is simply deduced $V_f = (V_1 + V_5)/2$. Two other tips (namely 2 and 4) are connected and biased negatively relative to the central tip (tip 3) by a constant voltage $\Delta V=42$~V that is applied from a Li-Ion external battery. When $\Delta V$ is sufficiently large currents $I_2$ and $I_4$ reach ion saturation current $I_i = (I_2 + I_4)/2$. Tip 3 collects an electron current and the electron temperature is deduced from $V_f$ and $V_3$ as $T_e = (V_3 - V_f)/\ln(3)$ (note that the factor $\ln(2)$ for a usual triple probe is replaced here by $\ln(3)$ due to a current collection on tip 3 which is twice the opposite of the ion saturation current). The plasma density is computed from both $I_{i}$ and $T_e$ using $n = I_i / 2 \pi R_p e c_s \alpha$, with $R_p$ the tips radius, $e$ the electron charge, $c_s = \sqrt{e T_e / M_i}$ the Bohm velocity and $\alpha$ a parameter depending on the ratio of the probe tip radius over the Debye length that accounts for the tips cylindrical geometry~\cite{Chen_2002}. The 0.8~mm tip spacing is much greater than the Debye length (15~$\mu$m): measurements can be considered independent. This spacing is of the order of the pixel resolution (0.7~mm/px): all tips of the probe will be considered at the same location when comparing the probe measurements to imaging.}

\N{The floating potential $V_f$ is measured from the voltage of the external tips 1 and 5 as $V_f = (V_1 + V_5)/2$. The plasma density and electron temperatures are deduced from the currents and voltage measurements of tips 1, 2 and 4. Tips 2 and 4 are biased negatively relative to the central tip 3 using a constant voltage $\Delta V=42$~V fed from an external Li-Ion battery. Currents $I_2$ and $I_4$ reach the ion saturation current $I_i = (I_2 + I_4)/2$ when the biasing $\Delta V$ is sufficiently large. Tip 3 collects an electron current (equal to $-2 I_i$) and the electron temperature is deduced from $V_f$ and $V_3$ as $T_e = (V_3 - V_f)/\ln(3)$ (note that the factor $\ln(2)$ for a usual triple probe is replaced here by $\ln(3)$ due to a current collection on tip 3 which is twice the opposite of the ion saturation current). The plasma density is computed from both $I_{i}$ and $T_e$ using $n = I_i / 2 \pi R_p e c_s \alpha$, with $R_p$ the tips radius, $e$ the electron charge, $c_s = \sqrt{e T_e / M_i}$ the Bohm velocity and $\alpha$ a parameter depending on the ratio of the probe tip radius over the Debye length that accounts for the tips cylindrical geometry~\cite{Chen_2002}. The electrical measurements from each tips can be confidently considered independent since their spacing (0.8~mm) is much larger than the Debye length ($\approx$15~$\mu$m). It is worth noting that the spacing between tips is of the order of the pixel resolution for the high-speed camera imaging (0.7~mm/px): all tips of the probe will thus be considered at the same location when comparing the probe measurements to imaging.}

\begin{figure}[h!]
    \centering
    \includegraphics[width = 0.98\columnwidth, trim={0in 0in 0in 0in},clip]{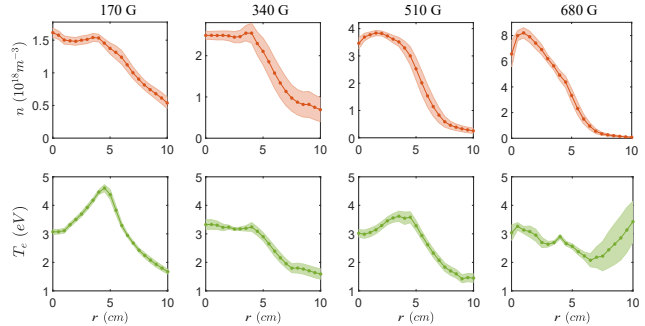}
    \caption{Radial profiles of the plasma density $n_e$ and electron temperature $T_e$, at $z=L_2$.}
    \label{n_Te_variations}
\end{figure}

Radial profiles of the plasma density and electron temperature are shown in Fig.~\ref{n_Te_variations} for values of the confining magnetic field ranging from $160$~G to $640$~G. The profiles of the standard deviation of temporal fluctuations are displayed as shaded colors. Time-averaged values and standard deviations were computed from 2 seconds measurements. The plasma density increases at the center of the plasma column, while it decreases at the edge of the plasma column, as the value of $B$ increases\footnote{ The local decrease of density at the center for $B=480$~G and $B=640$~G is expected to be due to the probe intrusiveness.}. The electron temperature peaks around $x=4.5$~cm at the lowest value of $B=160$~G, which can be understood as a signature of the inductive heating occurring over the penetration depth close to the source tube. This peak, although less marked, is still present up to the highest value of $B$. A last interesting feature is the strong increase of $T_e$ (both the time-averaged value and the amplitude of the fluctuations) at the edge of the column for $B=640$~G, which is well reproducible and not fully understood yet.

We underline that the probe provides simultaneous measurements of the temporal fluctuations of the plasma parameters. In the remainder of this article, light intensity recordings are systematically compared to models based on the dependence of emitted light intensity with $n$ and $T_e$, using collisional cross sections from the literature. Temporal fluctuations of light emission is also compared to the rough measurements of $I_i(t)$, $V_f(t)$ and $T_e(t)$ from the probe -- this choice of $I_i$ instead of $n$ was motivated by the fact that time series extracted from high speed camera imaging is usually compared to the ion saturation current fluctuations of Langmuir probe, used as a proxy for the plasma density fluctuations~\cite{Antar_2007, Oldenburger_2010, Brandt_2011, Cui_2015}.

\subsection{Visible light camera imaging}
\label{subsec:camera}

\begin{figure}
    \centering
    \includegraphics[width = 0.98\columnwidth, trim={0in 0in 0in 0in},clip]{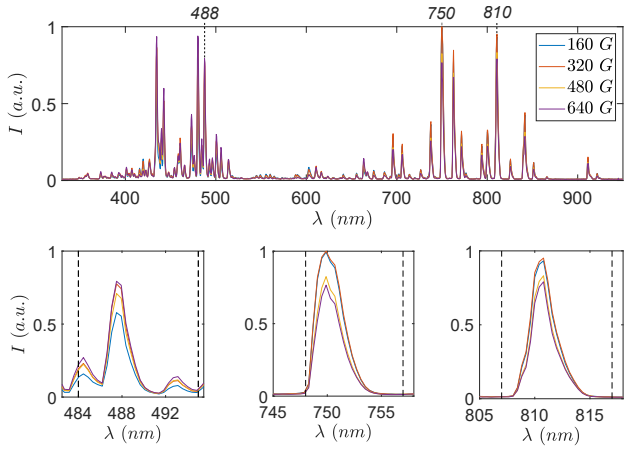}
    \caption{Spectra of the light naturally radiated by the plasma column at its center, for values of the magnetic field raging from $160$~G to $640$~G. Top : full spectra. Bottom : zoom in the value ranges of the optical filters that are used in this article for the camera measurements, around $488$~nm, $750$~nm and $810$~nm.}
    \label{spectral_lines}
\end{figure}

\N{Simultaneously to probe measurements, the light emitted by the plasma column is high-speed recorded.} The light radiated by the plasma is filtered using interferometric filters centered around three specific wavelengths, namely $488$~nm, $750$~nm and $810$~nm. Figure~\ref{spectral_lines} shows the spectra of the light emitted by the plasma column at its center, for increasing values of $B$, recorded using an OceanOptics USB 2000+ Spectrometer. The optical fiber collected the light along the $z$-axis (from the center of the optical borosilicate window). Peaks in light intensity are observed for each of the three central wavelengths of the filters, see bottom panel of Fig.~\ref{spectral_lines}, where the dark dashed lines show the limits of the Full Half Width Maximum transmission of the filters as provided by the manufacturer.
The electronic transitions at the origin of these spectral lines are provided in Table~\ref{table_levels_up_down}. The line emitted at $488$~nm is an ArII line, and we will neglect the contribution of the light emitted around $484$~nm when analyzing the images filtered around $488$~nm. All other transitions are ArI lines. The knowledge of the upper levels of the radiated light is used later on for the determination of the dependence of the light intensity with $n$ and $T_e$. Note finally that as the magnetic field increases, the intensity of the ArII line slightly increases whereas it \REM{increases} \N{decreases} for the neutral lines. 

\begin{table}[h!]
\renewcommand{\arraystretch}{1.5}
\scriptsize 
\begin{center}
 \begin{tabular}{c | c c c | c c c | c } 
 Line (nm) & \multicolumn{3}{c|}{Upper level} & \multicolumn{3}{c|}{Lower level} & Species \\  [0.5ex] 
 \hline
 487.99 &
 $3s^2 3p^4(^3\textrm{P})4p $&$ ^2\textrm{D}^{\circ} $&$ 5/2$ &
 $3s^2 3p^4(^3\textrm{P})4s $&$ ^2\textrm{P} $&$ 3/2$ &
 ArII  \\ [0.5ex] 
 \hline
 750.39 &
 $3s^2 3p^5(^2 \textrm{P}^0_{1/2})4p $&$ ^2[1/2] $&$ 0$ &
 $3s^2 3p^5(^2 \textrm{P}^0_{1/2})4s $&$ ^2[1/2]^{\circ} $&$ 1$ &
 \multirow{2}{*}{ArI} \\
 751.47 &
 $3s^2 3p^5(^2 \textrm{P}^0_{3/2}) 4p $&$ ^2 [1/2] $&$ 0$ &
 $3s^2 3p^5(^2 \textrm{P}^0_{3/2}) 4s $&$ ^2 [3/2]^{\circ} $&$ 1$  \\ [0.5ex]
 \hline
 810.36 &
 $3s^2 3p^5 (^2 \textrm{P}^{0}_{3/2}) 4p $&$ ^2[3/2]	$&$ 1 $ &
 $3s^2 3p^5(^{2}\textrm{P}^{0}_{3/2})4s $&$ ^{2}[3/2]^{\circ} $&$ 1$ & \multirow{2}{*}{ArI} \\	
 811.53 &
 $3s^2 3p^5 (^2 \textrm{P}^{0}_{3/2})4p $&$ ^2[5/2] $&$ 3 $ &
 $3s^2 3p^5 (^2 \textrm{P}^{0}_{3/2})4s $&$ ^2[3/2]^{\circ} $&$ 2 $    \\
\end{tabular}
\end{center}
\caption{Upper and lower energy levels of the dominant spontaneous radiated light transitions \N{analyzed in this article}, from the NIST database~\cite{NIST_2020}.}
\label{table_levels_up_down}
\end{table}

\REM{Filtered lines are} \N{Filtered light is} recorded using a Phantom v2511 camera facing the $z=80$~cm transparent chamber end. The central line of sight corresponds to the $z$-axis. In order to minimize the parallax effect, the distance between the camera and the chamber window is artificially increased by a set of two mirrors, reaching $3.5$~m in the present configuration. In spite of this arrangement, parallax cannot be neglected: a $10$~cm in diameter circle at the back of the chamber is seen on the images $ 22 \%$ smaller than a $10$~cm in diameter circle at the front of the chamber. The images are recorded at a frame rate of $200$~kfps, with an exposure time of $4.5$~$\mu s$, and by time series of $100$~ms. The camera trigger was also acquired on the data acquisition module, in order to get simultaneous measurements with the probe. The spatial resolution is $256 \times 256$~px$^2$ for a field of view of approximately $20$~cm, i.e. capturing fully the inside of the chamber. The focus is set at the position $z=L_2$. However the depth of field $DoF$ with the optical lenses used here (aperture f/4, focal length $135$~mm) is of the order of the chamber length \REM{(as a lower limit, by considering a very strict circle of confusion of the pixel resolution size, we already have $DoF \approx 54$~cm)} \N{(considering a very strict circle of confusion of the pixel resolution size as a lower limit, gives $DoF \approx 54$~cm)}. Hence we have to consider that the recorded light is the result of an integration over $z$, the implication of which are detailed in the next subsection.

\begin{figure}
    \centering
    \hfill
    \includegraphics[width = 0.478\columnwidth, trim={0in 0in 0in 0in},clip]{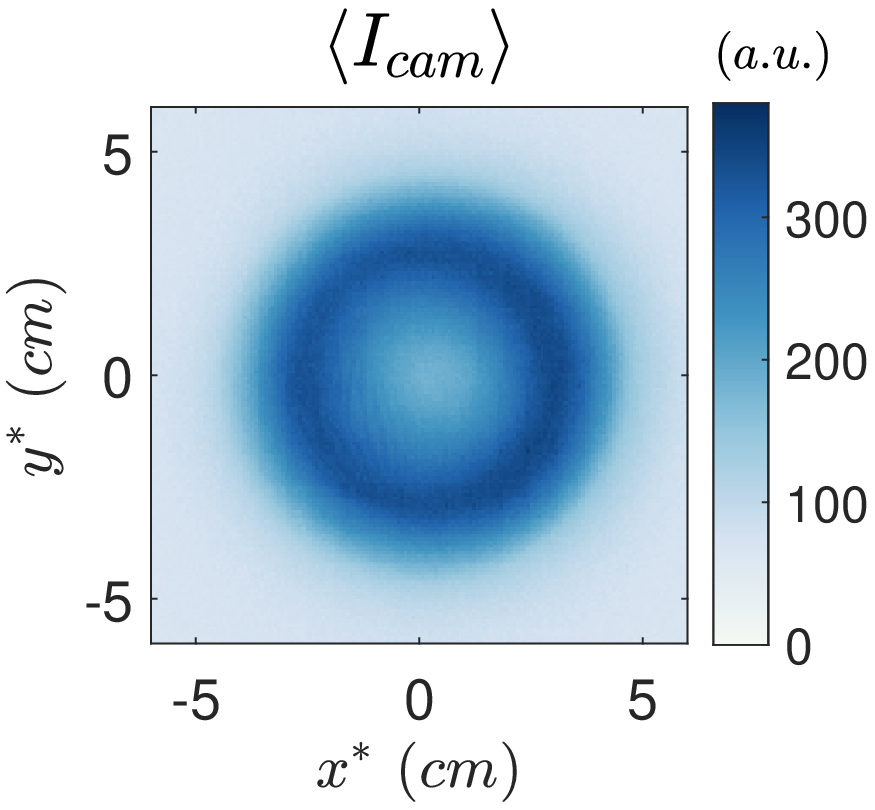}
    \hfill
    \includegraphics[width = 0.46\columnwidth, trim={0in 0in 0in 0in},clip]{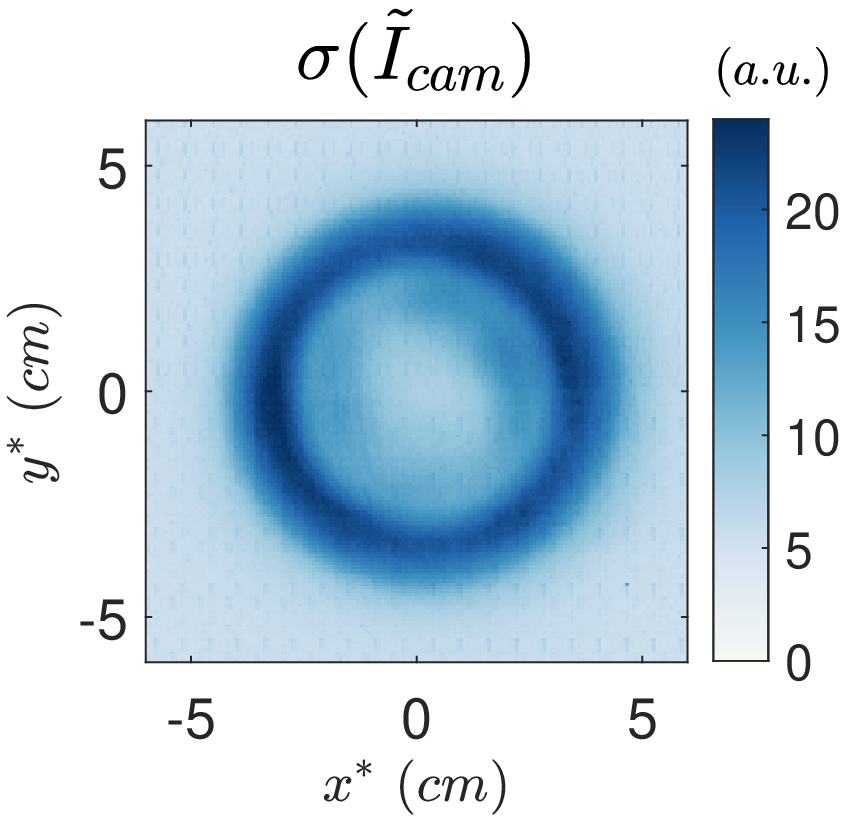}
    \hfill
    \caption{Images of the light recorded by the camera at $B=160$~G, filtered around $750$~nm: time averaged intensity (left) and the intensity fluctuations standard deviation (right).}
    \label{images_moy}
\end{figure}

The resulting image of the mean intensity $I_{cam}$ and the fluctuations standard deviation $\sigma(\tilde{I}_{cam})$ are plotted in Fig.~\ref{images_moy}, for the lowest value of the magnetic field ($B=160$~G) and using the $750$~nm filter. \N{Note that the camera images are displayed in a frame $(x^*,y^*)$, whose definition is given below. The first observation is that, as expected, the emitted light is axisymmetric. We thus discuss the features according to the radius $r^* = \sqrt{x^{* 2}+y^{* 2}}$ in the image plane.} The largest intensity is observed around a radius of $\sim 2.8$~cm, with a sharp gradient outward, which is maximal at $\sim 4.5$~cm. The amplitude of the fluctuations is of the order of 10~$\%$ of the total amplitude. The fluctuation pattern also presents a ring of large amplitude around $\sim 3.5$~cm. Similar features are observed on the images recorded with the other filters around $488$~nm and $810$~nm (see appendix~\ref{appendix:camera_profiles}). Surprisingly the light intensity profile shape \REM{is also roughly conserved with all} \N{only slightly evolves with the} magnetic field \N{amplitude} \REM{, up to $B=640$~G,} whereas the plasma parameters ($n$, $T_e$, $V_p$) radial profiles shapes \REM{all} vary \REM{with $B$} \N{as the magnetic field amplitude increases (see Fig.~\ref{n_Te_variations}}). This discrepancy in the radial profiles evolution between camera imaging and probe measurements is addressed in more details in section~\ref{sec:mean}. 
Note finally that the constant light intensity level for $r\gtrsim 5$~cm, visible on radial mean profiles displayed in appendix~\ref{appendix:camera_profiles}, does not correspond to light directly emitted by the plasma, which density and temperature strongly decrease at the column edge. \REM{This light is thought to be} \N{Light collected at the edge is likely} due to reflections on the cylindrical chamber and BN end, and therefore constitutes noise of the camera imaging data. This noise level, computed over the radial locations $r^*\leq5$~cm and the four magnetic field values, corresponds on average to $\sim 50$~$\%$ of the signal for the light intensity data filtered around 488~nm, $\sim 25$~$\%$ for 750~nm and $\sim 35$~$\%$ for 810~nm. For the comparison with probe profiles, $95$~$\%$ of the noise level is subtracted to each mean profile of the camera imaging data \N{in order to keep positive values for the light intensity}. An example of a resulting profile of $I_{cam}$ is shown in Fig.~\ref{n_Te_imaged_profiles_200A}.

\subsection{Geometrical corrections}
\label{subsec:camera_corrections}

\REM{Now before going further} It is important to note that a direct comparison between probe and camera light intensity radial profiles is not relevant. While probe measurements are done at a precise location on the z-axis, $I_{cam}$ is the result from a light integration over z, and because of the parallax effect and the inhomogeneity of the axial magnetic field the camera lines of sight scan locations with different values of the plasma parameters. In order make the data comparison possible, a transformation is applied to the probe profiles, \REM{that artificially reproduces} \N{modeling} the light integration process along the z-axis.

\begin{figure}[h!]
    \centering
    \includegraphics[width = 0.98\columnwidth, trim={0in 0in 0in 0in},clip]{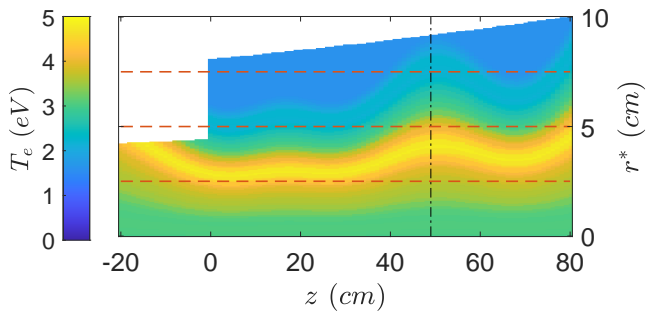}
    \caption{Spatial map of the electron temperature, reconstructed from the five tips probe profile performed at $z=L_2$ (dark dot-dashed line). The spatial domain inside the cylindrical chamber and the source is distorted so as to make the camera lines of sight parallel (red dashed lines), which makes the integration along $z$ easier. $B=160$~G.}
    \label{Te_map}
\end{figure}

The first assumption is that of cylindrical symmetry of the plasma column, \REM{which is justified from} \N{supported by} the images shown in Fig.~\ref{images_moy}. All the following computation are therefore performed in the 2D plane ($z, r>0$). The magnetic field spatial configuration is deduced from a numerical computation \N{taking into account the real geometry of the coils}, and is combined with the parallax effect computed from calibration images, to deduce the magnetic lines configuration as seen from the camera. 
The integration along the $z$-axis also requires a second assumption: frozen plasma parameters along magnetic field lines. This was checked by comparing radial profiles of $n$ and $T_e$ measured at $z=L_1$ and $z=L_2$, while the radial axis was transformed following the computed magnetic field lines, for the four set of $B \in [160; 640]$~G. The results show a very reasonable match with a discrepancy of the profiles shape along the radius of less than $\sim 20\%$ for both $n$ and $T_e$.
Under the last assumption, the plasma parameters for the full plasma column can be reconstructed from a single radial scan with the probe. An illustration of this process is given in Fig.~\ref{Te_map}, which shows a two-dimensional map of the electron temperature, reconstructed  from a profile measured at $z=L_2$ for $B=160$~G. Note that the map is plotted with respect to ($z, r^*$), the variables of the distorted space in which the camera lines of sight are parallel.
\REM{The axial variation of $n$ and $T_e$ of $\sim50$~$\%$ between $z=L_1$ and $z=L_2$ was not measured more precisely, and is not considered in this map reconstruction.}

\begin{figure}[h!]
    \centering
    \includegraphics[width = 0.7\columnwidth, trim={0in 0in 0in 0in},clip]{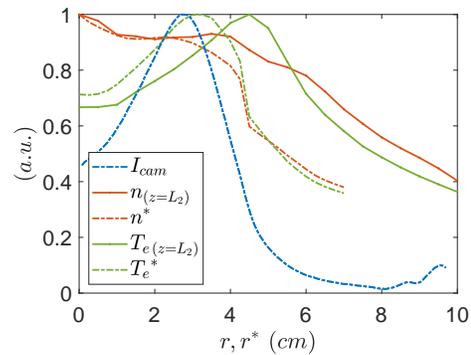}
    \caption{Radial profiles of light intensity $I_{cam}$ recorded by camera with the $750$~nm filter, to be compared to the density $n$ and the electron temperature $T_e$ measured by probe at $z=L_2$, as well as to their "image" profiles $n^*$ and $T_e^*$, for $B=160$~G.}
    \label{n_Te_imaged_profiles_200A}
\end{figure}

Integrated quantities may then be computed from the reconstructed \N{time-averaged} plasma parameters profiles in the $(r^*,z)$ plane as $ X^*(r^*) = \int_z X(r^*,z) dz$. The results of this process are shown in Fig.~\ref{n_Te_imaged_profiles_200A}, for $n^*$ and $T_e^*$ at $B=160$~G. \N{Figure~\ref{n_Te_imaged_profiles_200A} shows the profiles of the plasma density $n$ and electron temperature $T_e$ measured as a function of $r$ in the $z=L_2$ plane, and the integrated profiles $n^*$ and $T_e^*$ as a function of $r^*$}. The integrated profiles $n^*$ and $T_e^*$ are computed only up to $r^*=6.9$~cm, which corresponds to the first camera line of sight away from the center, that crosses the limiting magnetic field line which intercepts the chamber wall at $z=L_2$ (see Fig.~\ref{Te_map}). For $r^*>6.9$~cm, the single probe measurements at $L_2$ is not enough to reconstruct the plasma parameters on all field lines crossing the lines of sight. For comparison a radial profile of $I_{cam}$ is plotted in Fig. \ref{n_Te_imaged_profiles_200A}. It is computed as the average along $\theta$ of the mean camera image. The profile of the electron temperature $T_e$ measured at $z=L_2$ is peaked at $r \approx 4.5$~cm, while $T_e^*$ is peaked at $r^* \approx 3$~cm, which falls very close to the peak observed for the light intensity at 750~nm: this suggests the importance of $T_e$ for light emission recorded by the camera. Moreover, it confirms that the time averaged light intensity $ I_{cam}$ is not simply proportional to the time averaged plasma density for our experimental conditions. Similar conclusions hold for lines at 488~nm and 810~nm.

\section{Time averaged light intensity strongly depends on $T_e$}
\label{sec:Te}

The dependence on $T_e$ has to be considered in order to understand the profile of $I_{cam}$. In the present plasma conditions of low density ($n \sim 10^{18}$~m$^{-3}$), low degree of ionization ($5 \textnormal{-} 20$~$\%$), ion temperature of the order of a few tenths of an eV and low pressure ($p_0 \sim 1$~mTorr) implying $T_e \gg T_i \gg T_n$, \REM{we can use the approximation} \N{allowing the application}  of the Corona model~\cite{Fantz_2006}. As a consequence the intensity naturally radiated at a given wavelength only depends on the electron density and on the ground state density on the one hand, and, on the other hand, on the excitation rate constant $K_{ex}$ from the ground state to the radiative state (given as the "upper" states in Table~\ref{table_levels_up_down}).

The light intensity for each of the transitions identified in Table~\ref{table_levels_up_down} are thus expected to be proportional either to the functional form $f_n = n \, K_{ex}$ for the ArI lines or the functional form $f_{n^2}= n^2 \, K_{ex}$ for the ArII line -  this point will be discussed further below.
For each of the excited state, the excitation rate constant $K_{ex}$ strongly depends on $T_e$~\footnote{
\, The rate constant $K_{ex} = \langle \sigma_{ex} v \rangle_v$ can be expressed as $K_{ex} = \left( \frac{m_e}{2 \pi e T_e} \right)^{3/2} \displaystyle  \int_0^{\infty} \sigma_{ex}(v) v e^{-\frac{m v^2}{2 k T_e}} 4 \pi v^2 dv$
and recast as $K_{ex} = \left( \frac{e}{2 \pi m_e T_e} \right)^{1/2} \int_0^{\infty} \sigma_{ex}(\varepsilon) \, \varepsilon \, e^{-\frac{\varepsilon}{T_e}} d\varepsilon$
}
, and a non uniform radial profile of $K_{ex}$ is expected from the density and electron temperature profiles shown in Fig.~\ref{n_Te_variations}.  Assuming a Maxwellian electron distribution function, $K_{ex}$ can be computed from the cross section $\sigma_{ex}$.
Finally, $K_{ex}(T_e)$ is fitted numerically in the range [1, 5] eV (corresponding to the temperature of our experimental conditions) using the functional form $	K_{ex}(T_e) = K_{ex}^0 e^{-\epsilon_{ex} / T_e}$ with $\varepsilon_{ex}$ a fitting parameter equivalent to an averaged excitation energy~\cite{Lieberman}.

\begin{figure}[h!]
    \centering
    \hfill
    \includegraphics[width = 0.51\columnwidth, trim={0in 0in 0in 0in},clip]{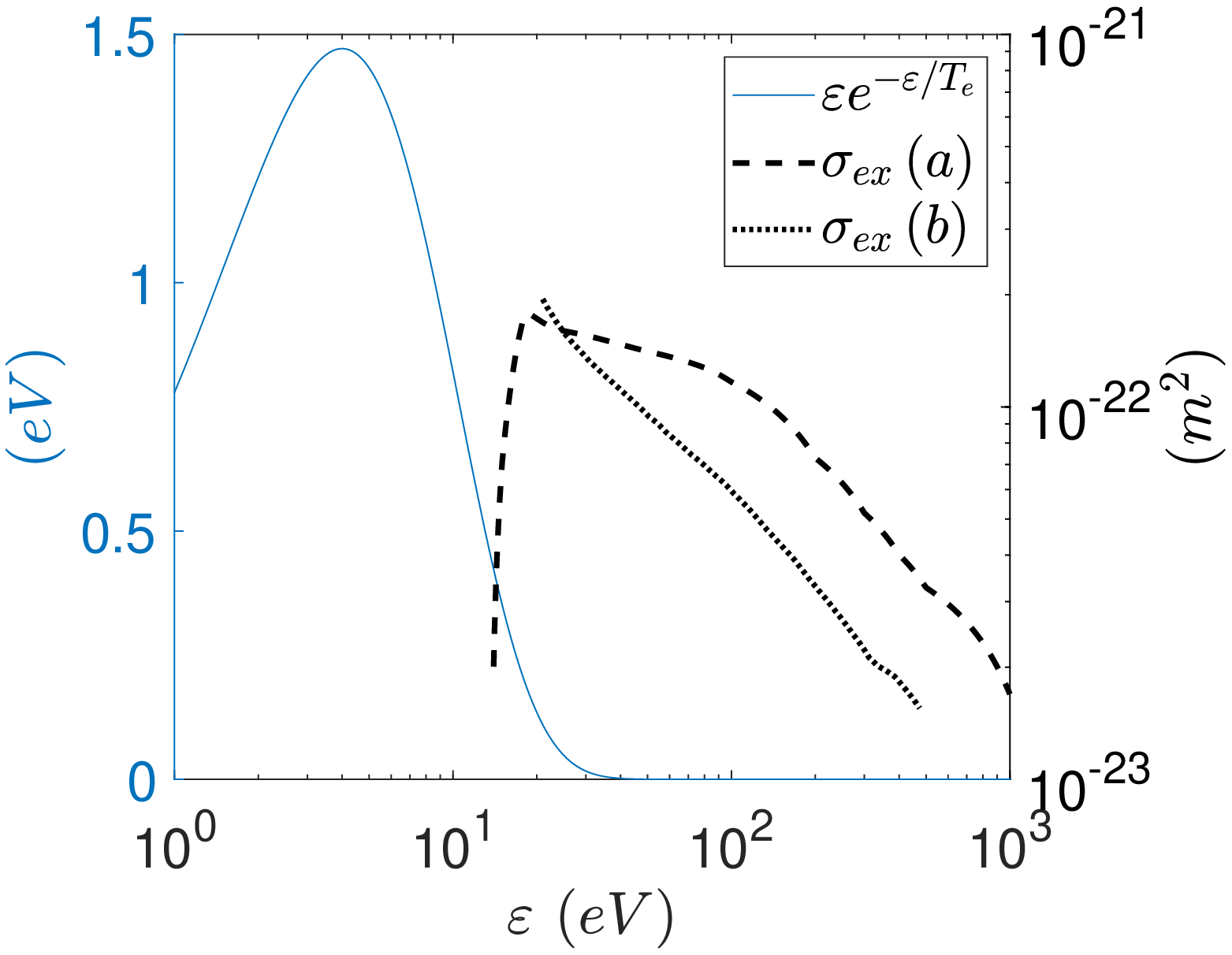}
    \hfill
    \includegraphics[width = 0.45\columnwidth, trim={0in 0in 0in 0in},clip]{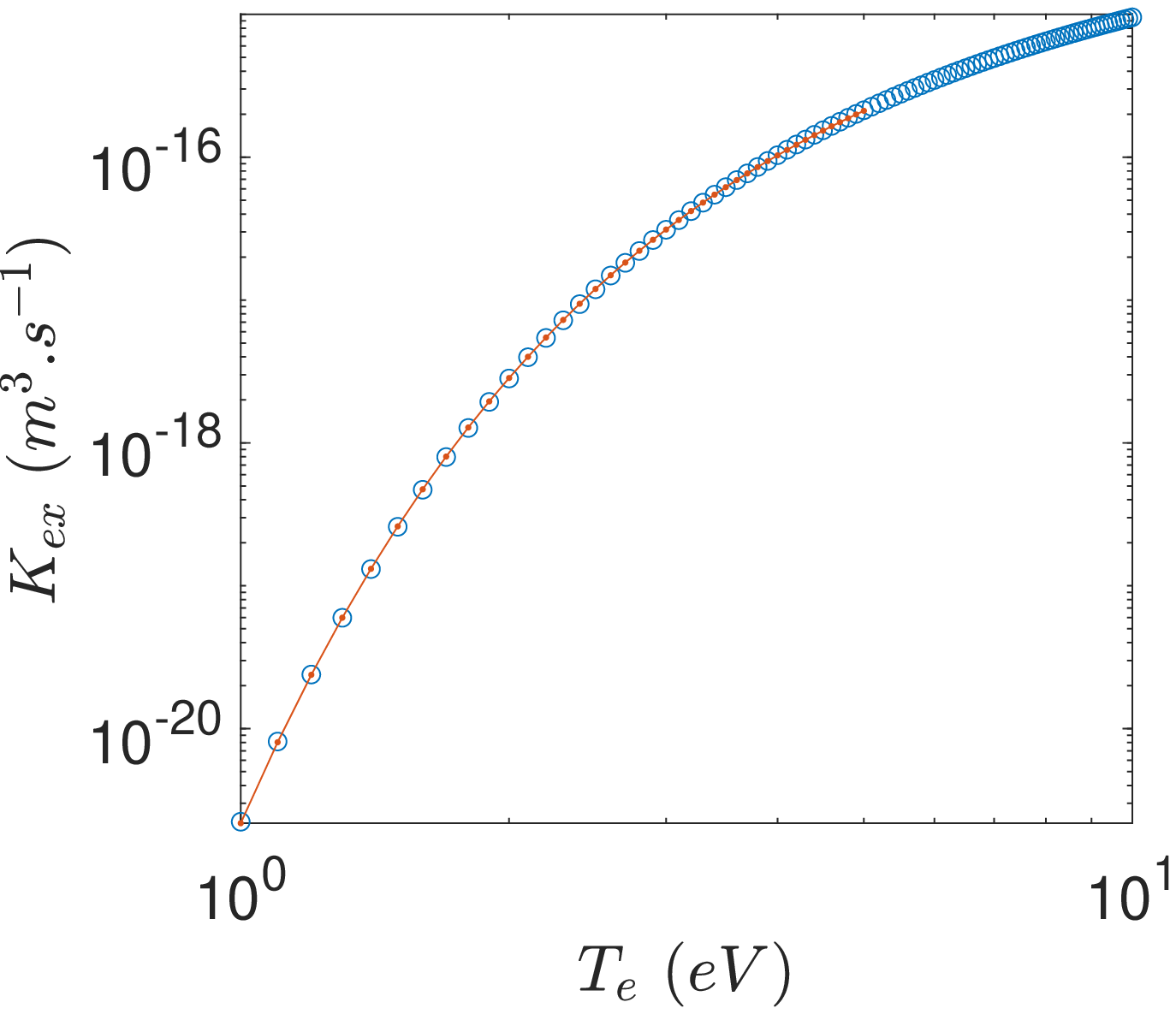}
    \hfill
    \caption{Left : Maxwellian distribution function example for $T_e = 4$~eV, plotted along with the excitation cross-sections of level $4p[1/2]0$ from Ar atom ground level (curve (a)) and of $4p ^2D^{\circ}_{5/2}$ from $\textrm{Ar}^+$ ground level (curve (b)). Right : computation of $K_{ex}(T_e)$ using $\sigma_{ex}^{(a)}$ (blue dots), and fit in the range $T_e \in [1 ; 5]$~eV with $f(T_e) = K_{ex,0} e^{-\varepsilon_{ex}/T_e}$  (red curve).}
    \label{K_ex_fit}
\end{figure}

Figure~\ref{K_ex_fit} (left) shows the cross sections of excitation processes $\textrm{Ar} \rightarrow \textrm{Ar}(4p[1/2]0)$ and  $\textrm{Ar}^+ \rightarrow \textrm{Ar}^+(4p \, ^2\textrm{D}^{\circ} \, 5/2) $ (the energy state ${Ar}^+(4p \, ^2\textrm{D}^{\circ} \, 5/2)$ will be denoted $Ar^{+ *}$ for simplicity in the following), along with an example for $T_e = 4$~eV of the integrand $\varepsilon e^{-\varepsilon/T_e}$.
\REM{Then an example of} The computed evolution of $K_{ex}(T_e)$ as a function of $T_e \in [1, 10]$~eV is plotted in Fig.~\ref{K_ex_fit} (right, open blue symbol) for the excitation process $\textrm{Ar} \rightarrow \textrm{Ar}(4p[1/2]0)$.
The values of $\sigma_{ex}(T_e)$ used for these computations are taken from the literature. For the ArI lines a recent and complete set of Argon cross sections was used~\cite{Alves_2014}.
Note that for the light filtered around $810$~nm, the average of the two equally contributing radiative states identified in Table~\ref{table_levels_up_down} is used to compute $K_{ex}$. \N{This is a rather strong assumption given that the line at 811.53 is known to be mostly populated from a metastable state and not from the ground state. However, we stress here that our goal is to demonstrate that the importance of the temperature fluctuations when interpreting high-speed camera imaging and that precise plasma spectroscopy is beyound the scope of the present article.}
While database providing cross sections for neutral lines are widely available~\cite{Alves_2014} the dataset for ArII lines at low electron temperature are very seldom in the literature.
The set of cross sections $\sigma_{ex}$ for the $Ar^+ \rightarrow Ar^{+ *}$ excitation process was extracted from recent simulations~\cite{Dipti_2016} and are shown in Fig.~\ref{K_ex_fit}a), dashed line. The average energy found for $\epsilon_{ex}^{Ar^+ \rightarrow Ar^{+ *}} = 20$~eV. 
Since the ground state for this excitation process is Argon ion, the resulting light intensity is expected to follow the functional form $f_{n^2}$.
A second excitation process has been identified in the literature for the $Ar^{+ *}$ state, from direct ionization and excitation from the neutral Ar atom ground level, and a second set of cross section was extracted from an experimental study~\cite{Strinic_2004}.
An average excitation energy $\epsilon_{ex}^{Ar\rightarrow Ar^{+ *}}=35.4$~eV is found, at the value close to the sum of the excitation energy $\epsilon_{ex}^{Ar^+\rightarrow Ar^{+ *}}$ and the 15.8 eV ionization energy of Argon atom. Note that the light intensity form this physical process is expected to follow the functional from $f_{n}$.

For all the major processes involved in populating the upper states referenced in~\ref{table_levels_up_down}, the results of the average excitation energies computed from the numerical fits of $K_{ex}$ \N{as a simple Arrhenius law $K_{ex}(T_e) = K_{ex}^0 e^{-\epsilon_{ex} / T_e}$}, as well as the functional form expected in terms of dependence with the density $n$, are summarized in Table~\ref{table_models}.
\REM{The form of Eq.~\eqref{eq::K_ex_fit_form}} \simon{The Arrhenius law} fits very well the rate constant dependencies in $T_e$, with a root mean square errors associated with the fits ranging from $0.7.10^{-2}$ to $3.8.10^{-2}$. All the fits are plotted in appendix~\ref{appendix:K_ex_fits}.

\begin{table}[h!]
\renewcommand{\arraystretch}{1.5}
\begin{center}
 \begin{tabular}{c c c c} 
 Filter (nm) & Exc. process & $\varepsilon_{ex}$ (eV) & $f$ \\ [0.5ex] 
 \hline
 \multirow{2}{*}{$488 \pm 5$} & $\textrm{Ar}^+ \rightarrow \textrm{Ar}^+(4p \, ^2\textrm{D}^{\circ} \, 5/2) $  & \textbf{20.0} &  $f_{n^2}$ \\
   & $\textrm{Ar} \rightarrow \textrm{Ar}^+(4p \, ^2\textrm{D}^{\circ} \, 5/2) $ &  35.4 &  $f_{n}$  \\ [0.5ex] 
  \hline
  $750 \pm 5$ & $\textrm{Ar} \rightarrow \textrm{Ar}(4p[1/2]0) $  & \textbf{14.4} &  $f_{n}$ \\  [0.5ex] 
  \hline
  \multirow{2}{*}{$810 \pm 5$} & $\textrm{Ar} \rightarrow \textrm{Ar}(4p[3/2]1) $   & \multirow{2}{*}{15.0} &  \multirow{2}{*}{$f_{n}$} \\
    & $\textrm{Ar} \rightarrow \textrm{Ar}(4p[5/2]3) $ &   &  
\end{tabular}
\end{center}
\caption{Excitation processes for light radiation at $488 \pm 5$~nm, $750\pm 5$~nm and $810\pm 5$~nm. The excitation rate was fitted as $K_{ex}(T_e) = K_{ex}^0 e^{-\varepsilon_{ex}/T_e}$ and the light intensity is expected to follow the functional form $f_{n}$ or $f_{n^2}$ (see text for details).}
\label{table_models}
\end{table}

\N{Note that the absolute spectroscopic calibration of the full optical acquisition chain is beyond the scope this work. Thus sections~\ref{sec:mean} and ~\ref{sec:fluctuations} focus on the comparison between normalized profiles of the light intensity and the simultaneously acquired plasma parameters. Section~\ref{sec:mean} addresses the comparison for time-averaged profiles, while section~\ref{sec:fluctuations} focuses on temporal fluctuations.  As a consequence, the comparison between normalized radial profiles allows to discriminate the accuracy of the following models :}
\begin{equation}
    \begin{aligned}
    f_n & = n e^{-\varepsilon_{ex}/T_e} \\
    f_{n^2} & = n^2 e^{-\varepsilon_{ex}/T_e} 
    \end{aligned}
    \label{eq:model_n_n2}
\end{equation}

\section{Mean profiles}\label{sec:mean}

\begin{figure}[h!]
    \centering
    \includegraphics[width = 0.98\columnwidth, trim={0in 0in 0in 0in},clip]{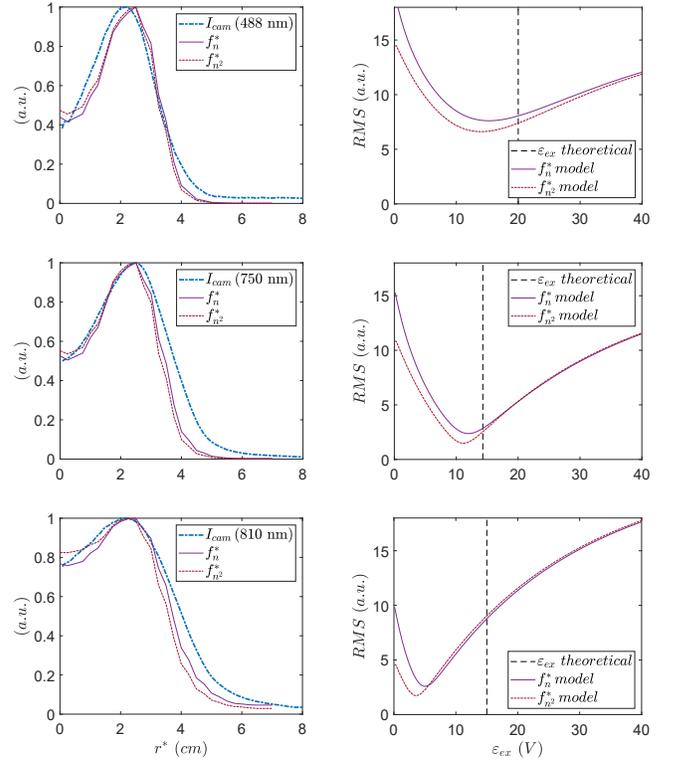}
    \caption{Comparison of radial profiles between the measure of $I_{cam}$, and its modelling from the plasma parameters using the models $f_n$ and $f_{n^2}$, for the filter at $488$~nm, $750$~nm and $810$~nm, at $B=320$~G. The respective optimization processes of $\varepsilon_{ex}$ (bottom) are plotted with the theoretical value coming from the numerical fit performed in section \ref{sec:Te}}
    \label{mean_profiles_Icam_model}
\end{figure}

In this section, the \N{time-averaged} experimental profiles recorded by the high-speed camera are compared to predictions using for the excitation rate $K_{ex}(T_e) \propto e^{-\epsilon_{ex} / T_e}$ 
and the \N{time-averaged} experimental radial profiles of the plasma parameters. Using the axisymmetric reconstruction of the plasma parameters in the $(r^*,z)$ plane introduced in subsection \ref{subsec:camera_corrections}, a time-averaged synthetic light radiation is computed as~:
\begin{equation}\label{eq:fstar}
    {f_{p}}^*(r^*, \varepsilon_{ex}) = \int_z f_{p}(n(r^*,z), \, T_e(r^*,z), \, \varepsilon_{ex}) dz,
\end{equation}
where $f_p$ stands for the functional dependency $f_n$ or $f_{n^2}$ introduced in the previous section, and, for clarity, the time averaged notation $\langle\cdot\rangle$ has been dropped. For each value of $\varepsilon_{ex}$ in the range $[0; 40]$~eV, the synthetic radial profile is compared to the profile of light intensity measured with the camera and averaged along $\theta$. The rms of the difference between the experimental and the synthetic profiles are displayed in the right panels of Fig.~\ref{mean_profiles_Icam_model} for $B=320$~G. The synthetic profiles which lower the difference with the experimental profiles are shown in the left panels of Fig.~\ref{mean_profiles_Icam_model}. For the sake of comparison both models $f_n^*$ and $f_{n^2}^*$ are systematically compared to the data.

The excitation energy that minimizes the error for the neutral $750$~nm line with the model $f_n^*$, of $\varepsilon_{ex}^{opt} = 12$~eV, is close to the theoretical value $14.4$~eV computed from the numerical fit (Table~ \ref{table_models}). The agreement is slightly lower for the $488$~nm line with the model $f_{n^2}^*$, with an optimal value of $\varepsilon_{ex}^{opt} = 14$~eV to be compared to $20.0$~eV from the numerical fit. Note that a very good match is also found with the $488$~nm data for a dependence of power law 1 in $n$ (model $f_{n}^*$), indicating that the parameter mostly responsible for the shape the radial profiles of light emission is the electron temperature through the  $e^{-\varepsilon_{ex}/T_e}$ term. As for the neutral $810$~nm line a very good match is found in the comparison with the model $f_n^*$, but for a value of $\varepsilon_{ex}^{opt} = 5$~eV three times lower than the one expected from the numerical fit. And again in this case no much difference is found between the models $f_n^*$ and  $f_{n^2}^*$ : the driving parameter for the reconstruction of $\langle I_{cam} \rangle$ is the average electron temperature.

As a result of the same analysis very satisfactory results of profile reconstruction (not shown here) are found for $B=160$~G. Note however that at this lower magnetic field the values of the average rate constant minimizing the error between $\langle I_{cam} \rangle$ and the model are relatively far from the theoretical ones computed in section \ref{sec:Te}, with $\varepsilon_{ex}^{opt} = 8.5$ and $\varepsilon_{ex}^{opt} = 3.9$ for $750$~nm and $810$~nm respectively with model $f_n$, and $\varepsilon_{ex}^{opt} = 11.6$ for $488$~nm respectively with model $f_{n^2}$. 

Finally the comparison is performed \REM{with the data measured at higher magnetic fields} at $B=480$~G and $B=640$~G.
\REM{There} \REM{, for which}
\simon{At these higher magnetic fields} the models do not match well anymore the light intensity profiles from the camera imaging. This is due to the systematic decrease of light intensity observed at the center of the plasma column, that cannot be reproduced by the $T_e$ profiles at these magnetic field values (see profiles for $B\geq480$~G in Fig. \ref{n_Te_variations}). \REM{The reason for this incompatibility between $\langle I_{cam} \rangle$ and $f_{p}^*$ profiles at high values of $B$ is still not fully understood. One possible explanation is the difference along the axis of the level of plasma radial diffusion and thermalisation. The peak of light intensity around $r^* \sim 2.5$~cm could then be explained by a peak in $T_e$ only present in the region $z\leq 0$ inside the source. Since this probable peak cannot be captured by probe measurements inside the chamber, the map reconstruction is likely to be inexact inside the source for $B \geq 480$~G.}\N{The discrepancy between the time-averaged emitted light and the models at high values of $B$ could be the signature of a peak in the electron temperature close to the source radius, in the source region, which is not measured in the electron temperature profiles measured at $z=L_2$. }

\section{Temporal fluctuations}
\label{sec:fluctuations}

\N{We now address the comparison between simultaneous measurements achieved with the five-tips probe and the camera, an important task for the interpretation of high-speed imaging to resolve the} \REM{dynamical behavior of plasma} \simon{plasma dynamical behavior.} \REM{In addition to the time averaged profiles discussed in the previous subsection, simultaneous measurements were achieved with the five-tips probe and the camera, for the same values of the magnetic field.}
The probe was inserted along the $x$-axis and the results presented here were acquired at a location $x_p=4$~cm of the probe tip (at $z=L_2$).
But due to the magnetic field ripple and the optical parallax, as previously discussed, the radial location $r$ at which the probe samples the signal corresponds to a location $r^{*}$ on the camera images closer to the center $r^{*} < r$.
The transformation process described in subsection~\ref{subsec:camera_corrections} is applied to a test radial profile composed of a single peak at $r=4$~cm and it is found that the peak is mostly translated at $r^*\sim 2.6$~cm.
Hence, even if this transformation is not reversible, the best area on the camera images to account for what is measured at $(z = L_2, x_p=4$~cm) is found around $x_p^* = 2.6$~cm .
The light intensity was therefore taken as the average of a $10 \times 10$ pixels box around $x_p^* = 2.6$~cm \N{and the simultaneous acquisition lasts 40 ms.} 
\REM{The probe measures last 1 second and the camera recording that is done with the same sample rate, only $40$~ms : the signal comparisons are made over the latter time length.}

\begin{figure}[h!]
    \centering
    \includegraphics[width = \columnwidth, trim={0in 0in 0.8in 0in},clip]{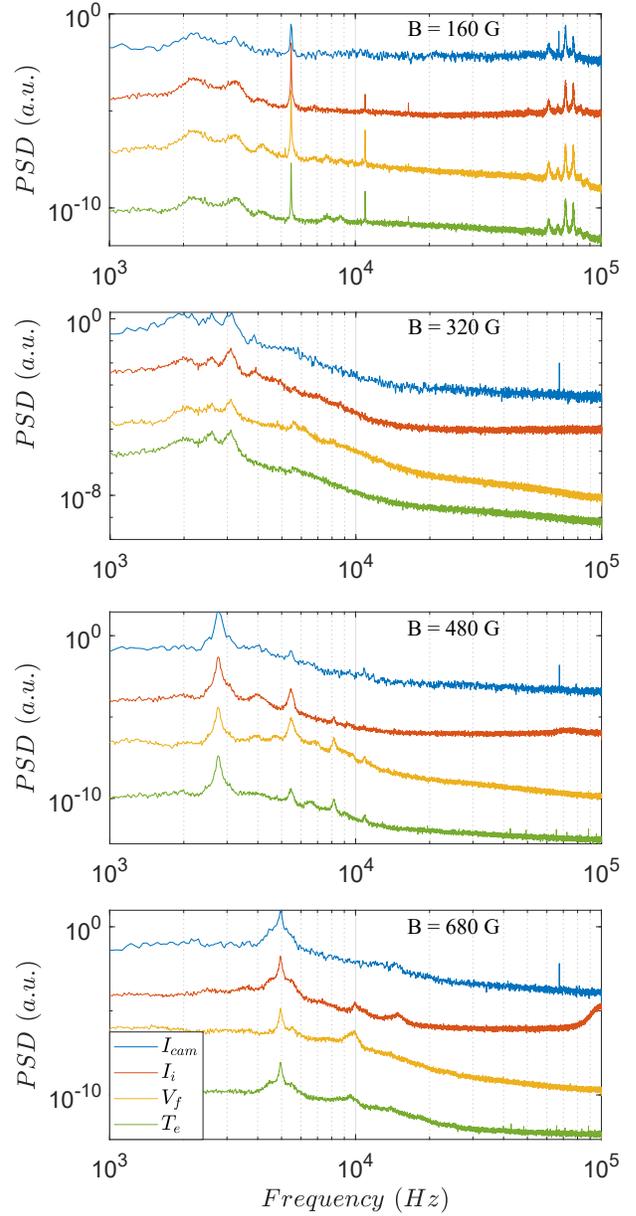}
    \caption{Spectra of the simultaneous camera ($I_{cam}$) and probe ($I_i$, $V_f$, $T_e$) measurements, with the $750$~nm optical filter, for $B= [160; 320; 480; 640]$~G.}
    \label{spectra_750nm}
\end{figure}

We recall here that the probe gives access to the instantaneous ion saturation current $I_i(t)$, from which the instantaneous plasma density $n(t)$ is computed as  $n(t)\propto I_i(t)/ \sqrt{T_e(t)}$. While the plasma density appears explicitly as a control parameter for the excitation processes, in this section we chose to present direct comparisons between $I_i$ and the light intensity, in order to provide a similar analysis to the results published in the literature using the ion saturation current from Langmuir probes as a proxy to the density. Note that, for the regimes reported in this study, the temperature fluctuations are a few percent, and the difference between the normalized fluctuations of $I_i$ and $n$  is of the order of $0.5 \, \tilde{T}_e / \langle T_e \rangle$ and can be most of the time neglected. The spectra of all the simultaneously measured signals are shown in Fig.~\ref{spectra_750nm}.

Except for the small peak at around 67 kHz for $B= [320, 480, 640]$~G, all the spectral features of the light intensity fluctuations correspond to features observed in the fluctuations of the ion saturation current, the electron temperature or the floating potential. At low magnetic field ($B=160$~G) the spectra are peaked around 70 kHz, and show very similar features for all the parameters. The waves observed at this low value of the magnetic field correspond to ion acoustic waves propagating azimuthally at the Bohm speed, and on the outer edge of the plasma column. For higher values of $B$, fluctuations are observed in the kHz range, and correspond once again to waves propagating in the azimuthal direction. These waves share features with unstable Kelvin-Helmholtz modes. The spectra for the fluctuations of the light intensity $I_{cam}$ are very similar to the spectra of the ion saturation current $I_i$. \REM{ : at $3.9$~kHz for $B=320$~G and $14.9$~kHz for $B=640$~A, peaks are observed on the spectra of $I_{cam}$ and $I_i$, whereas they are absent on the spectra of $V_f$ and $T_e$. Inversely the spectral components at $8.2$~kHz and $10.9$~kHz that are visible at $B=480$~G for $V_f$ and $T_e$, are much weaker and almost not noticeable for $I_{cam}$ and $I_i$.} However this rough comparison in the frequency domain is not sufficient to state which parameter controls the fluctuations of $I_{cam}$ fluctuations and a further analysis is detailed below.

\begin{figure}[h!]
    \centering
    \includegraphics[width = 0.98\columnwidth, trim={0in 0in 0in 0in},clip]{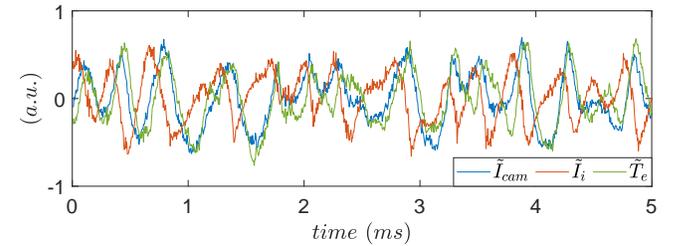}
    \caption{Simultaneous measurement of $\tilde{I}_{cam}$ with a filter at $750$~nm and $\tilde{I}_i$, $\tilde{T}_e$ by probe, for $B=320$~G.}
    \label{time_signals_200A_1}
\end{figure}

\begin{figure}[h!]
    \centering
    \includegraphics[width = 0.7\columnwidth, trim={0in 0in 0in 0in},clip]{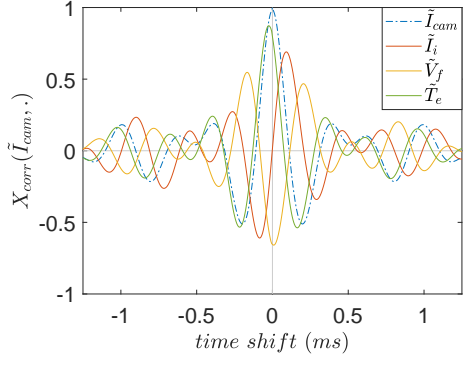}
    \caption{Correlation between the simultaneous measure of $\tilde{I}_{cam}$ with a filter at $750$~nm and $\tilde{I}_i$, $\tilde{T}_e$, $\tilde{V}_f$, for $B=320$~G.}
    \label{xcorr_signals_200A_1}
\end{figure}

Figure~\ref{time_signals_200A_1} shows simultaneous time series of the \N{normalized } fluctuations of $I_i$,  $T_e$ and $I_{cam}$ (filtered around $750$~nm) for $I_B=320$~G.  The correlations between $\tilde{I}_{cam}$ and the plasma parameters fluctuations, as well as $\tilde{I}_{cam}$ autocorrelation, are displayed in Fig.~\ref{xcorr_signals_200A_1}. From the time series, a strong correlation is observed between $\tilde{T}_e$ and $\tilde{I}_{cam}$, which is confirmed by the high value of the maximum correlation  $\max(X_{corr}(\tilde{I}_{cam}, \tilde{T_e})) = 0.87$. The correlation between the light intensity and the ion saturation current is lower, with $\max(X_{corr}(\tilde{I}_{cam}, \tilde{I_i})) = 0.68$ (though this is already a significant value, which is comparable to what was reported in similar plasma conditions~\cite{Oldenburger_2010, Cui_2015}). 
However, previous studies~\cite{Oldenburger_2010, Cui_2015} reported measurements for which $\tilde{I}_{cam}$ and $\tilde{I}_i$ were observed to evolve in phase, whereas we observe here a significant time-lag between $\tilde{I}_{cam}$ and $\tilde{I}_i$. On the contrary, here, no delay is observed between $\tilde{I}_{cam}$ and $\tilde{T}_{e}$. Finally for the sake of readability, the time series of $\tilde{V}_f$ are not shown, but the correlation with $\tilde{I}_{cam}$ shows a delay around $\pi$ (Fig.~\ref{xcorr_signals_200A_1}). This anti-correlation between $\tilde{I}_{cam}$ and $\tilde{V}_f$ confirms an observation that was recently made~\cite{Light_2019} for both the emission lines at $750$~nm and $488$~nm.



\begin{figure}[h!]
    \centering
    \includegraphics[width = 0.98\columnwidth, trim={0in 0in 0in 0in},clip]{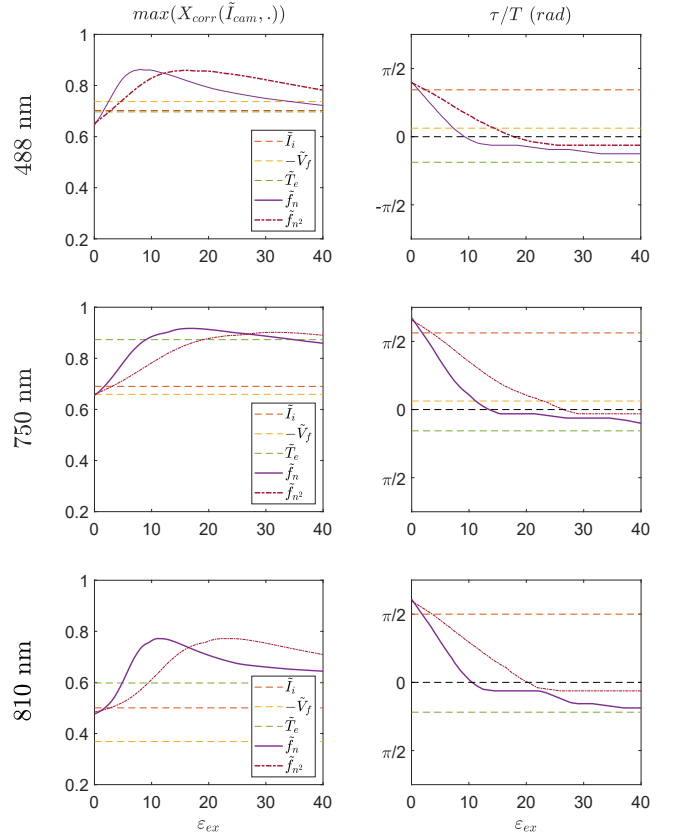}
    \caption{Maximal values (top) and associated time shifts (bottom) of the correlations computed between the time fluctuations signals of the light intensity $I_{cam}$ on one hand, and $I_i$, $V_f$, $T_e$ measured by probe, and the models $f_{n}$ and $f_{n^2}$ on the other hand, with $\varepsilon_{ex}$ in $[0; 40]$~eV. The results are presented for the optical lines around 488 nm, 750 nm and 810 nm, and for $B = 320$~G. The time shifts are normalized to the period $T$ of the main spectral component of $I_{cam}$.}
    \label{xcorr_optim_200A_1}
\end{figure}

We now compare in a systematic way the fluctuations of $I_{cam}$ with both models $f_n$ and $f_{n^2}$. Similarly to the study reported for the time-averaged profiles, for each value of the excitation energy $\epsilon_{ex}$ in the range $[0; 40]$~eV, synthetic  fluctuating series are computed according to the models (using the times series of the plasma parameters measured with the probe) and compared to $\tilde{I}_{cam}$. The results are shown in Fig.~\ref{xcorr_optim_200A_1} for the same data set as presented in Fig.~\ref{time_signals_200A_1} and Fig.~\ref{xcorr_signals_200A_1}, and for the three filters.

For the $750$~nm line, we observe an increased correlation with $\tilde{f}_{n}$, reaching 0.91 for an average excitation energy of $\varepsilon_{ex} = 17 \pm 3$~eV, close to the value computed in Table~\ref{table_models} $\varepsilon_{ex} = 14.4$~eV. It is also interesting to note that the maximum of correlation between $\tilde{I}_{cam}$ and $\tilde{f}_{n}$ corresponds to a zero time-delay, strengthening the validity of the results. For the sake of comparison, the other model $f_{n^2}$ is also tested : the correlation between $\tilde{I}_{cam}$ and $\tilde{f}_{n^2}$ is displayed in Fig.~\ref{xcorr_optim_200A_1}. A correlation up to 0.90, also associated with a zero delay between the signal, is observed for $\varepsilon_{ex} = 32 \pm 3$~eV. Hence with this model the correlation values are also increased with respect to that obtained with mere plasma parameters, but the optimal value found for $\varepsilon_{ex}$ is far from the reference value in Table~\ref{table_models}. This shows that, unlike with the mean profiles comparison presented in section \ref{sec:mean}, it is here essential to include the correct dependence in $n$ to recover the expected theoretical value of $\varepsilon_{ex}$.

Similarly, the time series of light fluctuations $I_{cam}$, measured at $488$~nm, are compared to the models. The maximum of correlation with the $f_{n^2}$ model, which reaches 0.85, is observed for $\varepsilon_{ex} = 17.5 \pm 4.5$~eV, close to the  value of $\varepsilon_{ex} = 20.0$~eV reported in Table~\ref{table_models}. This shows that the model $f_{n^2}$ matches very well the light intensity at 488~nm. The maximum of correlation with the $f_{n}$ model is observed for  $\varepsilon_{ex} = 8 \pm 2$~eV, which is far below $\varepsilon_{ex} = 35.4$~eV for the direct ionization plus excitation from the neutral Ar atom, presented in Table~\ref{table_models}. This result rules out the probability of this process for light emission at 488~nm.

Finally the results for the light emission at $810$~nm show that using both models the fluctuations correlation increases and the phase shift cancels out for the optimal $\varepsilon$ values. We get $\max(X_{corr})> 0.75$ for $\varepsilon_{ex} = 11.5 \pm 3$~eV with model $f_n$, and $\max(X_{corr})> 0.75$ for $\varepsilon_{ex} = 23 \pm 5$~eV with $f_{n^2}$. Even if the conclusion here cannot be as unequivocal as with the two other lines at 750 nm and 488 nm, the model $f_n$ that is expected to better match the data at 810 nm provides indeed a closer result to the theoretical value of $\varepsilon_{ex} = 15$~eV (Table~\ref{table_models}).

\begin{figure}[h!]
    \centering
    \includegraphics[width = 0.98\columnwidth, trim={0in 0in 0in 0in},clip]{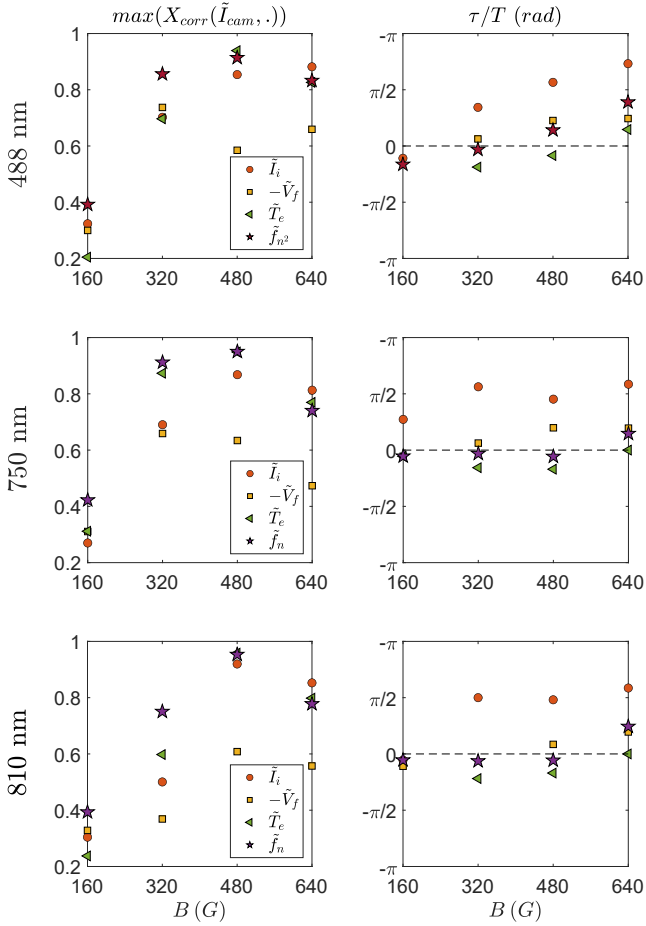}
    \caption{Summary of the maximal values (left) and associated time-shifts (right) of the correlations between the temporal fluctuations of $I_{cam}$ and $I_i$, $V_f$, $T_e$, $f_{p}(I_i, T_e, \varepsilon_{ex})$, for the 488 nm line ($\varepsilon_{ex} = 20.0$~eV  and $f_p = f_{n^2}$), the $750$~nm line (with $\varepsilon_{ex} = 14.4$~eV and $f_p = f_n$) and the $810$~nm line ($\varepsilon_{ex} = 15.0$~eV  and $f_p = f_{n}$), as a function of the magnetic field $B$. For each value of B, the time shifts are normalized to the period $T$ of the main spectral component of $I_{cam}$.}
    \label{xcorr_all_B}
\end{figure}

The same comparison is performed as well for $B=160$~G, $B=480$~G, $B=640$~G, but it is not described in details here. A sum-up of maximum correlation values and associated time shifts between $\tilde{I}_{cam}$ and $\tilde{I}_{i}$, $-\tilde{V}_{f}$, $\tilde{T}_{e}$ and the most accurate model ($\tilde{f}_{n}$ for the neutral lines at 750 and 810 nm, $\tilde{f}_{n^2}$ for the 488 nm ion line) is presented in Fig.~\ref{xcorr_all_B}, for $B$ ranging from $160$~G to $640$~G. Here the average excitation energies used in the models are taken equal to their theoretical values computed in Table~\ref{table_models}. 
Among plasma parameters the higher correlation is observed with $\tilde {T}_e$ for $I_B \leq 480$~G. \REM{Then at $I_B=640$~G, $\tilde{I}_i$ becomes the best correlated quantity to $\tilde{I}_{cam}$, but not significantly more than $\tilde{T}_e$. We can also see that $\tilde{I}_{cam}$ is delayed from $\tilde{I}_i$ by quarter of a typical oscillation period, and very close to be in phase with $\tilde{T}_e$, in almost all cases.}
\N{For $B=640$~G, $\tilde{I}_{cam}$ is highly correlated with $\tilde{I}_i$ and $\tilde{T}_e$. We can also see that $\tilde{I}_{cam}$ is delayed from $\tilde{I}_i$ by $\pi$, and very close to be in phase with $\tilde{T}_e$, in almost all cases.}
As for the models, apart from $B=640$~G, they clearly improve the correlation and reduce the time shift with the light intensity fluctuations. However the gain with respect to the mere electron temperature is not so significant, and in the end one might be tempted to consider more simply $\tilde{I}_{cam}$ to be a proxy for $\tilde{T}_e$ as a first order approximation. This is discussed in more details in the next section.

\section{Discussion}\label{sec:discussion}

The features shown in Fig.~\ref{spectra_750nm} can be much better understood with the spatio-temporal visualization of the plasma provided by fast camera imaging. At low value of the magnetic filed (160~G), the spectral components at 61 kHz and 72 kHz are identified without ambiguity, thanks to camera imaging, as modes $m=5$, $m=6$ of IAW waves, and the peaks at 5 kHz, 67 kHz and 77 kHz as the result of the weak non-linear interactions between them. At higher magnetic fields, the spectra maxima visible around 3 kHz and 5 kHz correspond to modes $m=2$ and $m=3$ of low frequency waves of the type Kelvin-Helmholtz, Rayleigh-Taylor of Drift-Waves, such as observed typically in linear devices~\cite{Burin_2005, Brochard_2005}. These results will be reported in forthcoming publications. These examples show the importance of camera imaging diagnostic for the investigation of non-linear evolution of instabilities and waves in magnetized plasma columns. In this context it is essential to have a fine understanding of what the camera images represent.

The results reported in section~\ref{sec:fluctuations} show that the $f_n$ and $f_{n^2}$ models accurately link fluctuations of the plasma parameters with the fluctuations of light emission, and we now discuss our results from first order expansions of these models :
\begin{equation}
    \begin{aligned}
        f_n\, \textrm{ model} & : \ \ \frac{\tilde{I}_{cam}}{\langle I_{cam} \rangle}
        =
        \frac{\tilde{n}}{\langle n \rangle}
        +  \xi \frac{\tilde{T}_{e}}{\langle T_{e} \rangle}  \\
        f_{n^2}\, \textrm{ model} & : \ \ \frac{\tilde{I}_{cam}}{\langle I_{cam} \rangle}
        =
        2 \frac{\tilde{n}}{\langle n \rangle}
        +  \xi \frac{\tilde{T}_{e}}{\langle T_{e} \rangle}
        \label{eq_balance_n_Te}
    \end{aligned}
\end{equation}
with $\xi =  \varepsilon_{ex} / \langle T_{e} \rangle$.
These expressions show that the balance between the normalized density and electron temperature fluctuations on the normalized light intensity fluctuations is uniquely set by the ratio $\xi = \varepsilon_{ex} / \langle T_{e} \rangle$. It is also interesting to note that this balance therefore depends on the location when the electron temperature is inhomogeneous.
For an average electron temperature of the order of $3-4$~eV,  $\xi$ reaches $\sim 5$. Hence for normalized density and electron temperature fluctuations of comparable orders of magnitude, the $\xi$ term drives $\xi \tilde{T}_{e} / \langle T_{e} \rangle $ to be the dominant terms in Eq.~\eqref{eq_balance_n_Te}.
Towards the edge of the plasma column, where $\langle T_{e} \rangle$ strongly decreases for $B$ lower than 480~G (see Fig.~\ref{n_Te_variations}), $\xi$ reaches values of the order of 10. Thus, at the edge of the plasma column, even though density fluctuations are twice larger than temperature fluctuations, the light intensity fluctuations are mainly driven by the fluctuations of electron temperature.

\begin{figure}[h!]
    \centering
    \includegraphics[width = 0.8\columnwidth, trim={0in 0in 0in 0in},clip]{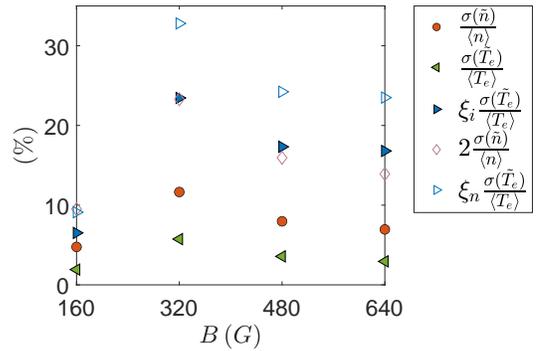}
    \caption{Comparison of the terms responsible for the light intensity fluctuations as expressed in Eq.~\eqref{eq_balance_n_Te}.}
    \label{xi_i_xi_n}
\end{figure}

Data from the simultaneous camera and probe measurements (at $r=4$~cm, see section~\ref{sec:fluctuations}) are analyzed in Fig.~\ref{xi_i_xi_n}, for increasing values of the magnetic field, where the amplitude of all the terms of the right hand sides of Eq.~\eqref{eq_balance_n_Te} are computed for all magnetic fields. \simon{The comparison are done here for the ion line at 750~nm and the neutral line at 488~nm.} \N{Note that $\xi_{i,n}$ are introduced as $\xi_{i,n} =  \varepsilon_{ex}^{i,n}/ \langle T_{e} \rangle $ with $\varepsilon_{ex}^{i} = 14.4$~eV and $\varepsilon_{ex}^{n} = 20$~eV for the ion and neutral lines respectively.}
\REM{Note that the comparison are done here using $\varepsilon_{ex} = 14.4$~eV for the $750$~nm line, and $\varepsilon_{ex} = 20$~eV for the $488$~nm line.} 
Though the plasma density exhibits higher normalized fluctuations (red dots) than the electron temperature (green triangles), the large value of $\xi$ leads to a dominant contribution of  $\xi \, \tilde{T}_{e} / \langle T_{e} \rangle $ in Eq.~\eqref{eq_balance_n_Te} (blue triangles). The term in $\tilde{T}_e$ is then twice larger than the term in $\tilde{n}$, for both lines and all $B$ field values : this explains well why $\tilde{I}_{cam}$ is globally better correlated to $\tilde{T_e}$ than to $\tilde{n}$ as found in section~\ref{sec:fluctuations}.

\begin{figure}[h!]
	\centering
    \includegraphics[width = 0.85\columnwidth, trim={0in 0in 0in 0in},clip]{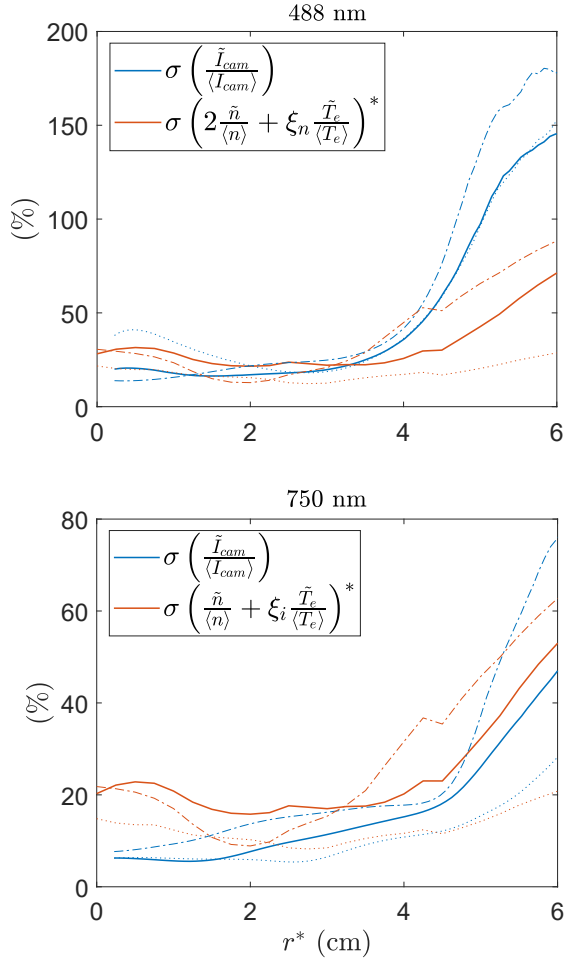}
    \caption{Comparison between the amplitudes of the normalized light intensity fluctuations, and its reconstruction with five-tips probe measurements of $n$ and $T_e$ in the range $r \in [0:6]$~cm, for wavelengths $488$~nm (left) and $750$~nm (right). $\xi_{i,n} =  \varepsilon_{ex}^{i,n}/ \langle T_{e} \rangle $ with $\varepsilon_{ex}^{i} = 14.4$~eV and $\varepsilon_{ex}^{n} = 20$~eV. The values of $B$ presented are $160$~G (dotted lines), $320$~G (full lines) and $480$~G (dash-dotted lines).}
    \label{I_cam_reconstruct}
\end{figure}

Let us now compare the amplitude of $\tilde{I}_{cam} / \langle I_{cam} \rangle$ to the prediction of Eq.~\eqref{eq_balance_n_Te}, in which the phase between $\tilde{n}$ and $\tilde{T}_e$ has to be taken into account. Using the five-tips probe measurements of $n$ and $T_e$, the quantity $\frac{\tilde{n}}{\langle n \rangle} + \xi \frac{\tilde{T}_{e}}{\langle T_{e} \rangle}$ is reconstructed for each value of the radius in the range $0 \leq r \leq 6$~cm, and its standard deviation is computed, yielding a radial profile along $r$. This profile is then transformed by the integration process described in subsection~\ref{subsec:camera_corrections} so as to be expressed along $r^*$. It is compared to radial profiles of the standard deviation of $\tilde{I}_{cam} / \langle I_{cam} \rangle$ in Fig.~\ref{I_cam_reconstruct}, with the light filtered at $488$~nm and $750$~nm, and $I_B = [160 ; 320 ; 480]$~G. Let us focus on the case $B = 320$~G (full lines in Fig.~\ref{I_cam_reconstruct}). \REM{At $488$~nm the match is very good inside the plasma column and relatively satisfactory at the edge. For $r \leq 4$ the normalized fluctuations of the light intensity are of $16 \%$ to $34 \%$, to be compared with reconstructed normalized fluctuations of $25 \%$ to $32 \%$. For $r \geq 4$ a discrepancy is observed : the normalized fluctuations get up to $145 \%$ at $r^* = 6$~cm for the light intensity, i.e. twice higher than that of the reconstructed signal of $71 \%$. 
At $750$~nm, the match is extremely satisfactory for $3.5 \leq r^* \leq 6$. The normalized fluctuations are very close between the light intensity and the reconstructed signal, with respective levels of $14 \%$ and $18 \%$ at $r^*=3.5$~cm, and of $47 \%$ and $53 \%$ at $r^*=6$~cm. Near the center the match is less good : at $r^*=1$~cm the standard deviation of $\tilde{I}_{cam} / \langle I_{cam} \rangle$ is of $6 \%$, and of $21 \%$ for the reconstructed signal.}
\N{At $488$~nm the match is very good inside the plasma column and relatively satisfactory at the edge. At $750$~nm, the match is extremely satisfactory for $3.5 \leq r^* \leq 6$, and slightly less towards the center.}
\REM{Overall the reconstruction process of the light intensity normalized fluctuations form probe measurements, using a first order approximation of the model given in Eq.~\eqref{eq:model_n_n2}, is satisfactory for $B = 320$~G. The amplitude of $\tilde{I}_{cam} / \langle I_{cam} \rangle$ can be recovered from the plasma parameters within a factor of $\sim 0.5$ to $\sim 3$. Similarly at the magnetic field values of $B = 160$~G (dotted lines in Fig.~\ref{I_cam_reconstruct}) and $B = 480$~G (dashed lines) reasonably good matching results are found.}
\N{Overall, the reconstruction process of the light intensity normalized fluctuations form probe measurements, using a first order approximation of the model given in Eq.~\eqref{eq:model_n_n2}, is satisfactory. The order of magnitude of $\tilde{I}_{cam} / \langle I_{cam} \rangle$ can be recovered from the plasma parameters.}

\section{Conclusion} \label{sec:conclusion}

Visible light naturally radiated by an Argon weakly magnetized plasma column was compared to plasma parameters acquired from an electrostatic probe. Using a Corona model, the origin of the light was interpreted in terms of simple excitation processes that were identified, and whose dependence on both the density and the electron temperature was modelled following a modified Arrhenius laws $n^a e^{-\varepsilon_{ex}/T_e}$ with $a=1$ for the neutral lines, and $a=2$ for the ion one. The average excitation energy $\varepsilon_{ex}$  was deduced from numerical fits, using cross sections from the literature. The model was then compared to camera recorded light for values of the magnetic field up to $640$~G. This was done for mean radial profiles, where geometrical corrections have to be taken into account, and for temporal fluctuations for which simultaneous measurements of camera and probe were performed.

More specifically, the following points were highlighted:
\begin{itemize}
\item The influence of electron temperature fluctuations cannot be ignored for a correct interpretation of the filtered light fluctuations, over a wide range of plasma parameters and diverse plasma fluctuations. 
\item The influence of the plasma density on light intensity (i.e. whether light emitted scales as $n$ or $n^2$) is nearly indistinguishable for the time-averaged profiles. Contrarily, it is very important for light fluctuations intensity.
\item For most of the reported regimes, the correlation between the electron temperature and the light fluctuations is larger than the correlation between the plasma density  and the light fluctuations. A first order approximation of the modified Arrhenius law shows that this is expected when the energy excitation greatly exceeds the electron temperature. It also gives an efficient estimation whether $\tilde{I}_{cam}$ should rather be considered as reflecting the density or electron temperature fluctuations, if not a combination of both.
\end{itemize}

Our systematic study demonstrated the importance of considering the influence of the electron temperature for the correct interpretation of high speed imaging of plasma emitted light, which cannot be simply considered as a proxy for the plasma density fluctuations.

\section{Acknowledgements}

This work was partly supported by French National Research Agency under contract ANR-
13-JS04-0003-01. We acknowledge support from the CNRS for the acquisition of the high-speed camera, and useful discussions with V. Désangles, G. Bousselin and E. De Giorgio.

\section*{Appendix A : Camera imaging radial profiles}
\label{appendix:camera_profiles}

\begin{figure}[h!]
	\centering
    \includegraphics[width = 0.98\columnwidth]{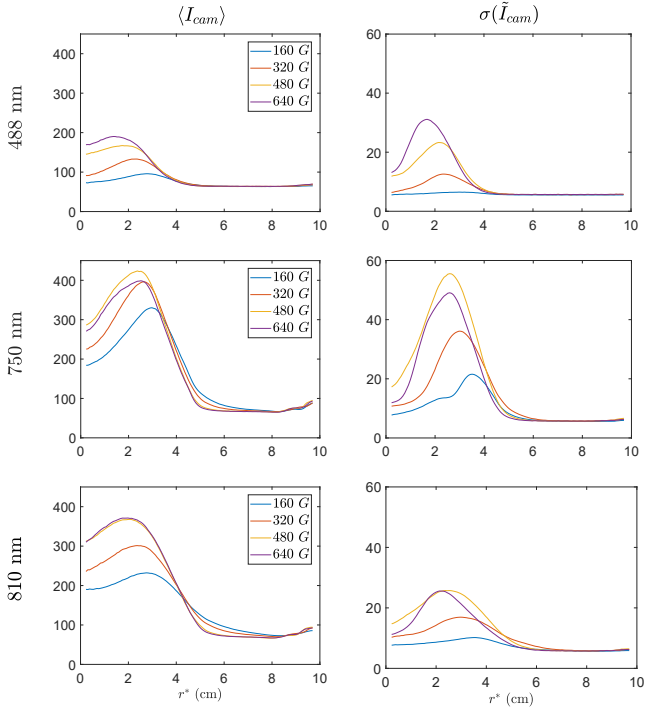}
	\caption{Radial profiles of the mean amplitude and fluctuations standard deviation of the light intensity, for $B= [160; 320; 480; 640]$~G.}
    \label{signal_noise}
\end{figure}

The radial profiles of the light intensity mean values and fluctuation amplitude are presented in Fig.~\ref{signal_noise}.
These profiles show a relatively poor signal/noise ratio for the filter around 488~nm, ranging from 1.2 at the plasma column center for $B=160$~G, to a value of 3 when the light collected is maximal around $r=1.5$~cm for $B=640$~G. For the filter at 750~nm this ratio is significantly improved, ranging from 2.7 ($r=0$~cm, $B=160$~G) to 6.2 ($r=2.5$~cm, $B=640$~G). At 810~nm the signal/noise ratio varies between 2.8 at the plasma center for $B=160$~G, to 5.4 at $r=2$~cm and $B=640$~G.

\section*{Appendix B : Fits of the rate constants}
\label{appendix:K_ex_fits}

\begin{figure}[h!]
	\centering
    \includegraphics[width = 0.9\columnwidth]{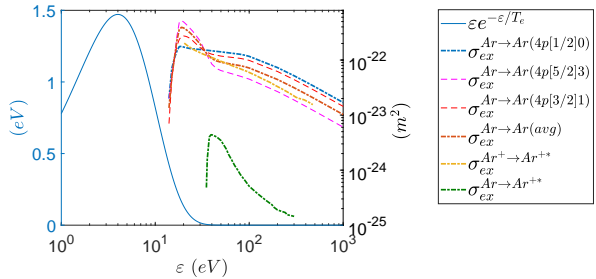}
	\caption{Cross sections of all the processes reported in table~\ref{table_models}, plotted along with an example of the $K_{ex}$ integrand $ \varepsilon e^{\varepsilon/T_e}$, for $T_e = 4$~eV.}
    \label{}
\end{figure}

\begin{figure}[h!]
	\centering
    \includegraphics[width = 0.45\columnwidth]{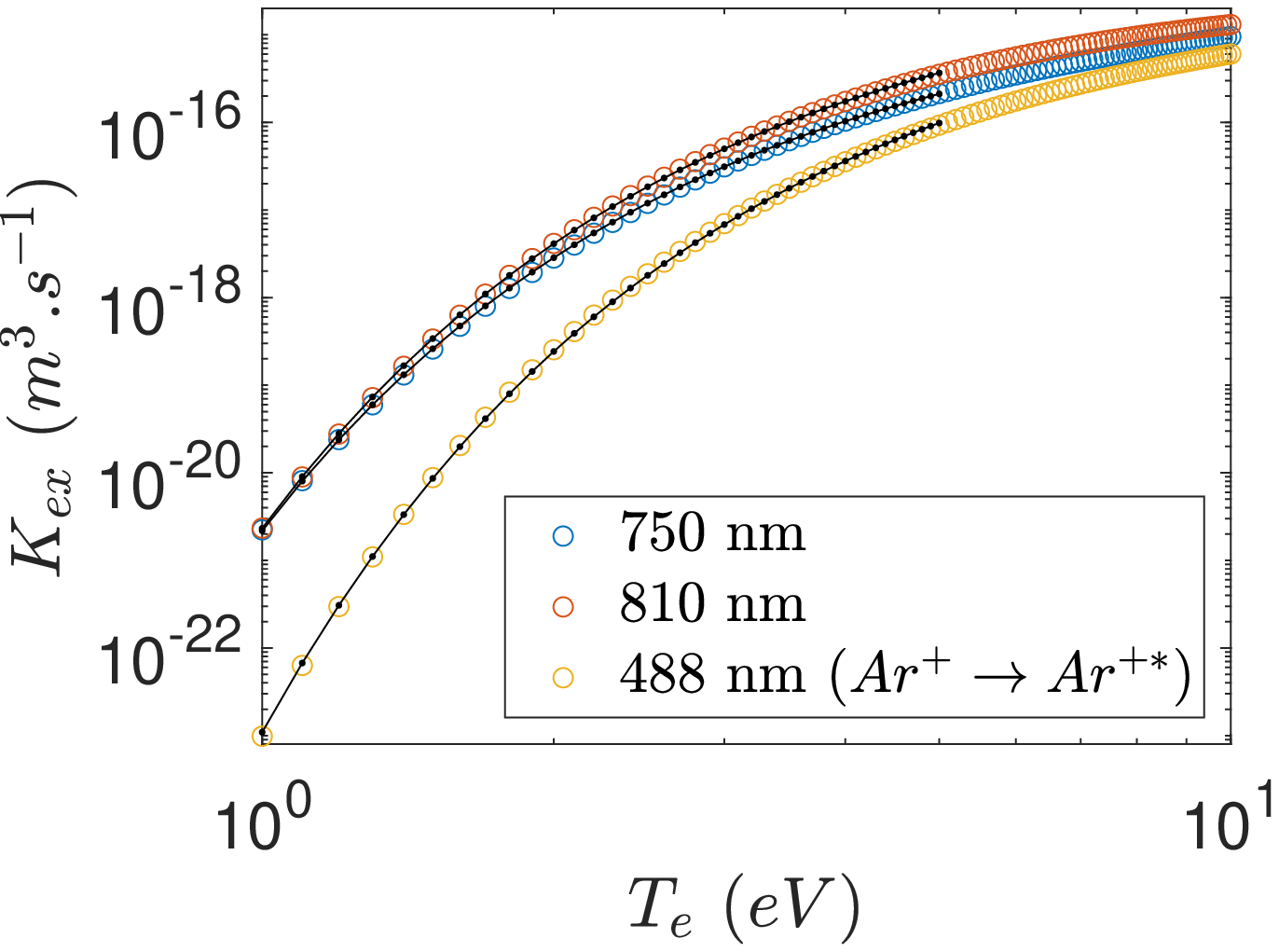}
    \includegraphics[width = 0.45\columnwidth]{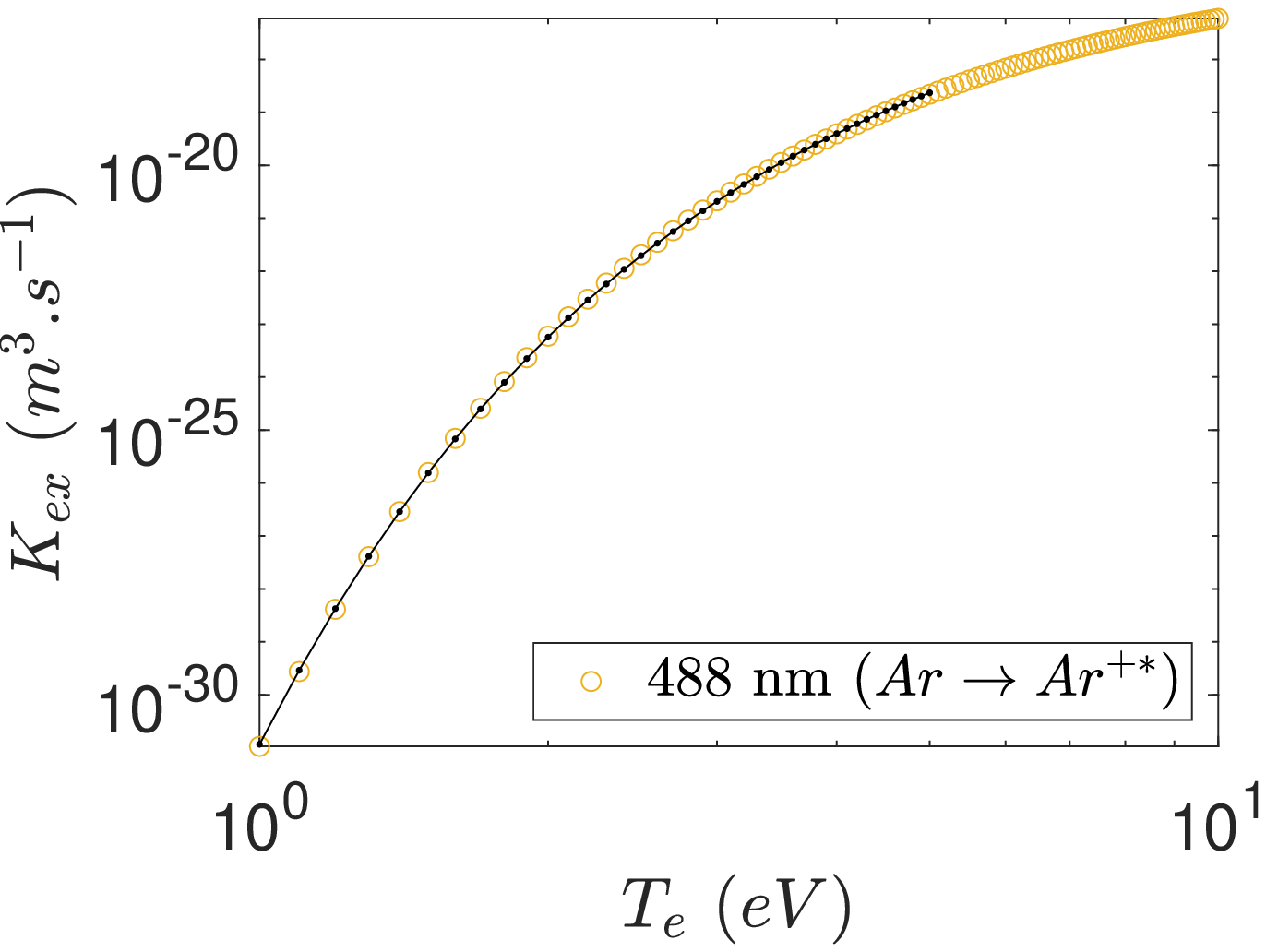}
	\caption{Fits of $K_{ex}(T_e)$ as a function of the form $K_{ex}^0 e^{-\varepsilon_{ex}/T_e}$, yielding the averaged excitation energies $\varepsilon_{ex}$ reported in table~\ref{table_models}.}
    \label{}
\end{figure}

\section*{Appendix C : Integration reverse process}
\label{appendix:integration_test}

\begin{figure}[h!]
    \centering
     \includegraphics[width = 0.75\columnwidth, trim={0in 0in 0in 0in},clip]{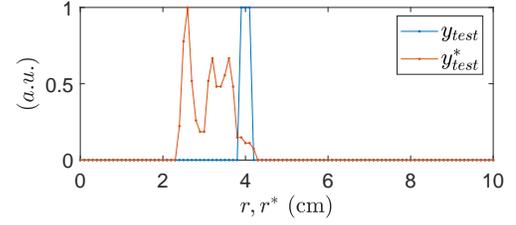}
    \caption{Test of the integration process described in subsection~\ref{subsec:camera_corrections}, applied to an Heaviside function centred around $r=4$~cm.}
    \label{integration_test}
\end{figure}

\section*{References}

%

\end{document}